\documentclass[12pt]{article}%
\usepackage{cite}
\usepackage{graphicx}
\usepackage{multicol}
\usepackage{amsfonts}
\usepackage{amssymb}
\usepackage{amsmath}
\usepackage{heck}
\usepackage{hyperref}
\usepackage{setspace}
\usepackage[all]{xy}
\usepackage{verbatim}
\usepackage{color}
\usepackage{cancel}%
\providecommand{\U}[1]{\protect\rule{.1in}{.1in}}
\numberwithin{equation}{section}

\newcommand{\Tr}{\, {\rm Tr}}

\begin{document}

\date{March 2011}

\preprint{MCTP-11-11}

\institution{IAS}{\centerline{${}^{1}$School of Natural Sciences, Institute for Advanced Study, Princeton, NJ 08540 USA}}

\institution{HARVARD}{\centerline{${}^2$Jefferson Physical Laboratory, Harvard University, Cambridge, MA 02138 USA}}

\institution{MICH}{\centerline{${}^3$Michigan Center for Theoretical Physics, University of Michigan, Ann Arbor, MI 48109 USA}}

\title{The Conformal Sector of F-theory GUTs}

\authors{Jonathan J. Heckman\worksat{\IAS}\footnote{e-mail: {\tt jheckman@ias.edu}}, Cumrun Vafa\worksat{\HARVARD}\footnote{e-mail: {\tt vafa@physics.harvard.edu}} and Brian Wecht\worksat{\MICH}\footnote{e-mail: {\tt bwecht@umich.edu}}}%

\abstract{D3-brane probes of exceptional
Yukawa points in F-theory GUTs are natural hidden sectors
for particle phenomenology. We find that coupling the probe to
the MSSM yields a new class of $\mathcal{N} = 1$ conformal fixed points
with computable infrared R-charges. Quite surprisingly, we find that
the MSSM only weakly mixes with the strongly coupled sector in
the sense that the MSSM fields pick up small exactly computable anomalous dimensions. Additionally, we
find that although the states of the probe sector transform as
complete GUT multiplets, their coupling to Standard Model fields
leads to a calculable threshold correction to the running of the
visible sector gauge couplings which \textit{improves}
precision unification.  We also briefly consider scenarios in which SUSY is broken in the hidden sector.
This leads to a gauge mediated spectrum for the gauginos and first two superpartner generations,
with additional contributions to the third generation superpartners and Higgs sector.}

\maketitle

\enlargethispage{\baselineskip}

\setcounter{tocdepth}{2}
\tableofcontents

\newpage

\section{Introduction}

F-theory provides a promising starting point for connecting string theory
with particle phenomenology. In particular, Grand Unified Theories (GUTs)
can be elegantly realized in this setup \cite{BHVI, BHVII, DWI, DWII} (see
\cite{Heckman:2010bq, Weigand:2010wm} for reviews). Such scenarios feature higher
dimensional GUT theories where the GUT group is localized on a seven-brane,
and matter is trapped at the intersection of the GUT brane with additional
flavor seven-branes. Interactions are localized at common intersections of the
matter fields. The F-theory approach to particle physics is promising because
while it is flexible enough to accommodate the basic features of realistic
particle phenomenology, it is also rigid enough to impose non-trivial
constraints on potential scenarios of physics  beyond the Standard Model (SM).

One example of a such a constraint is that generating
a large top quark mass requires the existence of
exceptional (e.g. E-type) symmetry enhancement points in the geometry
\cite{BHVI, DWI}. Another example is that in geometrically minimal
realizations of flavor hierarchies, both the CKM matrix and the mass matrices
are in relatively close accord with observation \cite{HVCKM, HVCP,
FGUTSNC, BHSV}. Certain supersymmetry (SUSY) breaking scenarios such as a ``PQ
deformation'' of minimal gauge mediation can also be accommodated
\cite{HVGMSB} in which the messenger fields localize on curves of the
geometry. Combining this with other geometric conditions
imposes constraints missing from purely low energy considerations; for example, in
many models the representation content of the messengers is fixed by the
available ways to unfold a singularity \cite{EPOINT}. This is
quite economical and constitutes a self-contained package. Nevertheless, given
the existence of an entire landscape of possible closed string vacua, it is
important to investigate possible well-motivated extensions of this framework.

A robust feature of this setup is that the possible ways to extend such GUT
models by additional seven-brane intersections are rather limited. Aside from
seven-branes, the main ingredients generically present in compactifications of
F-theory are probe D3-branes. Such D3-branes fill 3+1 noncompact spacetime
dimensions and sit at points of the internal geometry. Their presence in a globally complete
compactification is often required in order to cancel tadpoles. Moreover, they
are also locally attracted to the E-type Yukawa points of the visible sector
\cite{Funparticles}. It is therefore natural to ask what type of extra sectors
are realized on such probe theories. At a generic internal point of the
geometry, this extra sector is not terribly interesting. For example, a single
D3-brane realizes $U(1)$ $\mathcal{N}=4$ Super Yang-Mills theory. In this
theory, the vevs of the three chiral fields control the local position of the
D3-brane in the three complex-dimensional internal geometry.

However, there is a distinguished point in F-theory phenomenology where
intersecting seven-branes realize an $E_{6,7,8}$ exceptional symmetry. In geometrically
minimal constructions, a single point of $E_8$ is responsible for generating all
of the relevant visible sector Yukawa couplings \cite{BHSV, EPOINT}. Since the D3-brane is attracted to such Yukawa
points (see \cite{Martucci, FGUTSNC, Funparticles}), it is natural to
ask what happens if D3-branes reside at (or very near) such a point. It was
with this motivation in mind that these theories were proposed as possible
quasi-hidden sectors for F-theory GUTs in \cite{Funparticles} which were
further studied in \cite{FCFT, D3gen}.

The dominant couplings between the probe and Standard Model (SM) degrees of freedom
have been studied in \cite{Funparticles, TBRANES}. Some of the states of the
probe are charged under $SU(5)_{GUT}$ and therefore communicate via gauge
interactions with the visible sector. Another source of probe/MSSM couplings is
via F-terms. These F-terms primarily couple the probe to the Higgs fields and
third generation of matter fields. This is because in the flavor hierarchy
scenario of \cite{HVCKM, BHSV, EPOINT}, these are the modes with maximum
overlap with the Yukawa point, which is also the location of the D3-brane.

In the limit where one neglects the effects of the Standard Model fields, it
was found in \cite{FCFT} that such probe theories realize a new class of
strongly interacting $\mathcal{N} = 1$ superconformal field theories (SCFTs). These
theories are defined by starting from the $\mathcal{N} = 2$
Minahan-Nemeschansky theories with $E_8$ flavor symmetry \cite{MNI, MNII},
and adding a set of field-dependent mass deformations in the $SU(5)_{\bot}$
factor of $E_{8} \supset SU(5)_{GUT} \times SU(5)_{\bot}$.
The choice of mass deformations is dictated by the local configuration of
intersecting seven-branes. Remarkably, the configurations which lead to
new $\mathcal{N} = 1$ fixed points are precisely those which are of most
relevance for phenomenology.

With an eye towards potential model building applications, in this paper we
take the next step in this analysis by coupling the probe to a dynamical
visible sector. In other words, at the GUT scale $M_{GUT}$, we consider
performing a further deformation of the $\mathcal{N} = 1$ theories studied in
\cite{FCFT}. This deformation induces a flow to
another class of interacting CFTs. Much as in \cite{FCFT}, we can compute
various details of the system such as the scaling dimension of chiral
operators, as well as the running of the visible sector gauge couplings. Although the combined
probe/MSSM system realizes an interacting conformal fixed point, there is a
well-defined sense in which the mixing between the two sectors remains small.
Indeed, we find that the scaling dimensions of the SM chiral fields remains
rather close to their free field values, with anomalous dimensions typically in
the range of $0.0 - 0.1$.

Since the CFT living on the D3-brane includes degrees of freedom charged under
$SU(5)_{GUT}$, these modes must develop a mass to have evaded detection thus
far. This places a lower bound on the CFT breaking scale of order
$M_{\cancel{CFT}} \gtrsim$ TeV. It is natural to ask how to break conformal
invariance. A simple possibility is to have the D3-brane slightly displaced
from the Yukawa point, since the displacement automatically introduces a scale and gives mass to these
$SU(5)_{GUT}$ charged states of the probe sector.

\begin{figure}
[ptb]
\begin{center}
\includegraphics[
height=4.6198in,
width=6.1765in
]%
{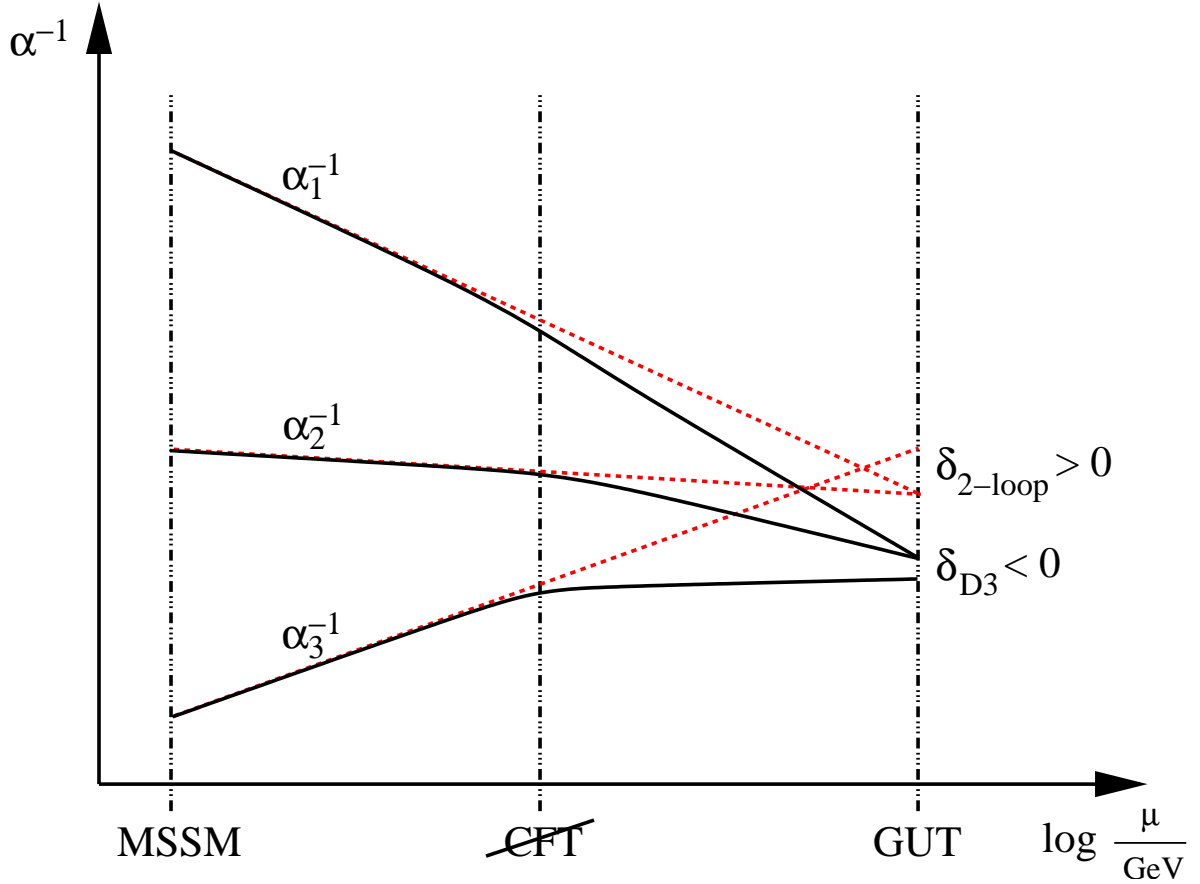}%
\caption{Depiction (not drawn to scale) of threshold corrections to gauge
coupling unification from the D3-brane probe sector (black solid curves). This
is to be compared with two-loop MSSM\ and other contributions with similar
effects (red dashed lines). The GUT\ scale is defined as the scale at which
$\alpha_{1}$ and $\alpha_{2}$ unify. The D3-brane threshold correction enters
at the scale of conformal symmetry breaking, which is the characteristic mass
scale for states of the probe charged under $SU(5)_{GUT}$. Here, we have
indicated that the size of the relative shift in couplings $\delta
\equiv(\alpha_{3}^{-1}-\alpha_{GUT}^{-1})/\alpha_{GUT}^{-1}$ is positive for
two-loop effects and negative for the D3-brane threshold. Balancing these
effects improves precision unification.}%
\label{unifyplot}%
\end{center}
\end{figure}

Treating the SM gauge group as a weakly gauged flavor symmetry of the conformal
fixed point, we calculate the contribution of the probe to the running of the SM gauge couplings.
Introducing a characteristic scale $M_{\cancel{CFT}}$ associated with conformal symmetry
breaking, we can view this computation as a threshold correction to the
running of the MSSM gauge couplings which enters at energies $M_{\cancel{CFT}}
<E < M_{GUT}$. We find that although the states of the probe sector fill out
full GUT multiplets, the additional charged degrees of freedom of the CFT
\textit{improve} precision unification of the MSSM (which for typical superpartner masses
otherwise contains an order $\sim 4\%$ mismatch at two loop level) due to their $SU(5)_{GUT}$
breaking couplings to the Higgs sector (see also \cite{Donkin:2010ta}). See
figure \ref{unifyplot} for a depiction of these effects.

It is also natural to ask what effect the D3-brane has on the visible sector at energies
below $M_{\cancel{CFT}}$. At this point there are two possibilities, depending on whether or not
SUSY breaking takes place on the D3-brane hidden sector. If it
does not, the story is rather simple, and the only remaining effects may
include some light remnants which interact via kinetic mixing with
the visible sector \cite{Funparticles, D3gen}.

If, however, SUSY breaking takes place on the D3-brane, the pattern of soft supersymmetry breaking terms in the visible sector may be strongly affected.
Along these lines, the most natural thing to assume is that the D3-brane position mode
which breaks conformal invariance also develops a SUSY breaking vev. Assuming this, the $SU(5)_{GUT}$ charged mediator states from
the probe to the visible sector play the role of messenger fields in a minimal
gauge mediation scenario, leading to a very predictive superpartner mass
spectrum. The dominant coupling to the third generation and Higgs fields
induces further corrections to the gauge mediated spectrum. These additional
contributions can induce additional third generation/Higgs sector
soft masses and large $A$-terms, as well as $\mu$ and $B \mu$-terms.

The organization of this paper is as follows. In section \ref{sec:QUASI} we
review the basic features of the visible sector in F-theory GUTs, and the
precise definition of the probe sectors we study in the remainder of this
paper. In section \ref{sec:FIXED} we study the infrared fixed points
associated with coupling the MSSM to the probe. Using these results, in
section \ref{sec:UNIFY} we demonstrate that for appropriate threshold scales,
the presence of the probe can actually improve precision unification. We next
turn in section \ref{sec:SUSY} to the potential role of the probe as a sector
for supersymmetry breaking and transmission to the visible sector. Section
\ref{sec:CONC} contains our conclusions. A brief review of probes of T-Branes
is given in Appendix A, and a collection of additional monodromy scenarios is collected
in Appendix B.

\section{Quasi-Hidden Sectors From F-theory}

\label{sec:QUASI}

In this section we review the basic setup we shall be considering. The main
idea is that the Standard Model is realized via a configuration of
intersecting seven-branes in F-theory, and that D3-branes constitute an
additional sector which can interact non-trivially with this system.
Throughout, we shall work in terms of a limit where four-dimensional \ gravity
has been decoupled, so that $M_{pl}^{(4d)}\rightarrow\infty$. This
approximation is especially well-justified in systems where we focus on just
the local intersections of non-compact seven-branes probed by a D3-brane which
is pointlike in the internal directions. We now turn to a more precise
characterization of the system we shall study.

\subsection{Intersecting Seven-Branes and the Visible Sector}

F-theory is a non-perturbative formulation of IIB\ string theory in which the
axio-dilaton $\tau=C_{0}+i/g_{s}$ can attain order one values, which moreover
can have non-trivial position dependence on the internal directions. Going to
strong coupling provides a way to realize intersecting brane configurations
with E-type symmetries. E-type symmetries are quite important in string-based
GUT\ models, thus suggesting F-theory GUTs as a natural framework for string
model building \cite{BHVI,BHVII,DWI,DWII}. The profile of the
dilaton is, in minimal Weierstrass form, dictated by the equation:
\begin{equation}
y^{2}=x^{3}+f(z,z_{1},z_{2})x+g(z,z_{1},z_{2}).
\end{equation}
Here, $z$, $z_1$ and $z_2$ define three local complex coordinates for the threefold base of
any F-theory compactification. The modular parameter of this elliptic curve is the axio-dilaton $\tau_{IIB}$.
The location of the intersecting seven-branes is then given by the
discriminant
\begin{equation}
\Delta\equiv4f^{3}+27g^{2}=0.
\end{equation}

In such models, the visible sector is realized on a configuration of
interecting seven-branes. The local intersections of such seven-branes can be
modelled in terms of eight-dimensional Super Yang-Mills theory with gauge
group $G$. We denote by $z_{1}$ and $z_{2}$ two local complex coordinates
parameterizing the internal worldvolume of the parent theory. The breaking
pattern of $G$ then dictates the locations of localized matter, as well as the
interactions between this matter \cite{KatzVafa, BHVI, DWIII, TBRANES}. For the purposes of
GUT\ model building, it is most convenient to work in terms of $E_{8}$ gauge
theory. The particular choice of a breaking pattern in a local patch of the
geometry is dictated by an adjoint-valued $(2,0)$ form which we denote by
$\Phi$. This $\Phi$, along with the internal gauge fields of the seven-branes,
satisfy a coupled system of F- and D-term equations. For example, to
realize an $SU(5)$ F-theory GUT, one considers the breaking pattern:%
\begin{equation}
E_{8}\supset SU(5)_{GUT}\times SU(5)_{\bot}%
\end{equation}
where $\Phi$ takes values in the adjoint of $SU(5)_{\bot}$. This
tilting of the seven-brane configuration occurs at an energy scale
which we shall refer to as $M_{\ast}$. The
compactification of the seven-brane is assumed to occur at the GUT\ scale
$M_{GUT}<M_{\ast}$. Numerically, we have $M_{\ast}\sim10^{17}$ GeV and
$M_{GUT}\sim2\times10^{16}$ GeV \cite{BHVII}.

The particular choice of $\Phi$ dictates the phenomenology of the visible
sector. A very convenient way to analyze supersymmetric configurations of
intersecting seven-branes is in terms of \textquotedblleft holomorphic
gauge.\textquotedblright\, This choice amounts to analyzing just the F-term equations
of motion for the system, and effectively complexifying the gauge group to
$G_{%
\mathbb{C}
}$. In an appropriate gauge, one can then locally present $\Phi(z_{1},z_{2})$
as a purely holomorphic expression in the local coordinates $z_{1}$, $z_{2}$
of the parent theory worldvolume coordinates. In realistic F-theory GUT
configurations, the actual seven-brane configuration corresponds to a
\textquotedblleft T-brane\textquotedblright\ \cite{TBRANES}. Locally, this configuration is specified
by the condition that at the origin, $\Phi(0,0)$ is
nilpotent but non-zero. In holomorphic gauge, $\Phi(0,0)$ can then be decomposed
into a direct sum of nilpotent Jordan blocks.

\subsection{Review of Probe D3-Branes}

So far, our discussion has focussed on the properties of the visible sector,
e.g., the seven-branes of the system. An additional well-motivated sector to
consider is one with D3-brane probes of such intersecting brane configurations. Away
from the configuration of seven-brane intersections, the probe D3-brane is
given by $\mathcal{N}=4$ Super Yang-Mills theory with $U(1)$ gauge group. The
worldvolume gauge coupling is the IIB axio-dilaton $\tau_{IIB}=\tau_{D3}$. As
the D3-brane moves closer to the configuration of intersecting seven-branes,
additional modes will become light. This can lead to highly non-trivial
interacting superconformal field theories.

\begin{figure}
[ptb]
\begin{center}
\includegraphics[
height=4.6224in,
width=4.9917in
]%
{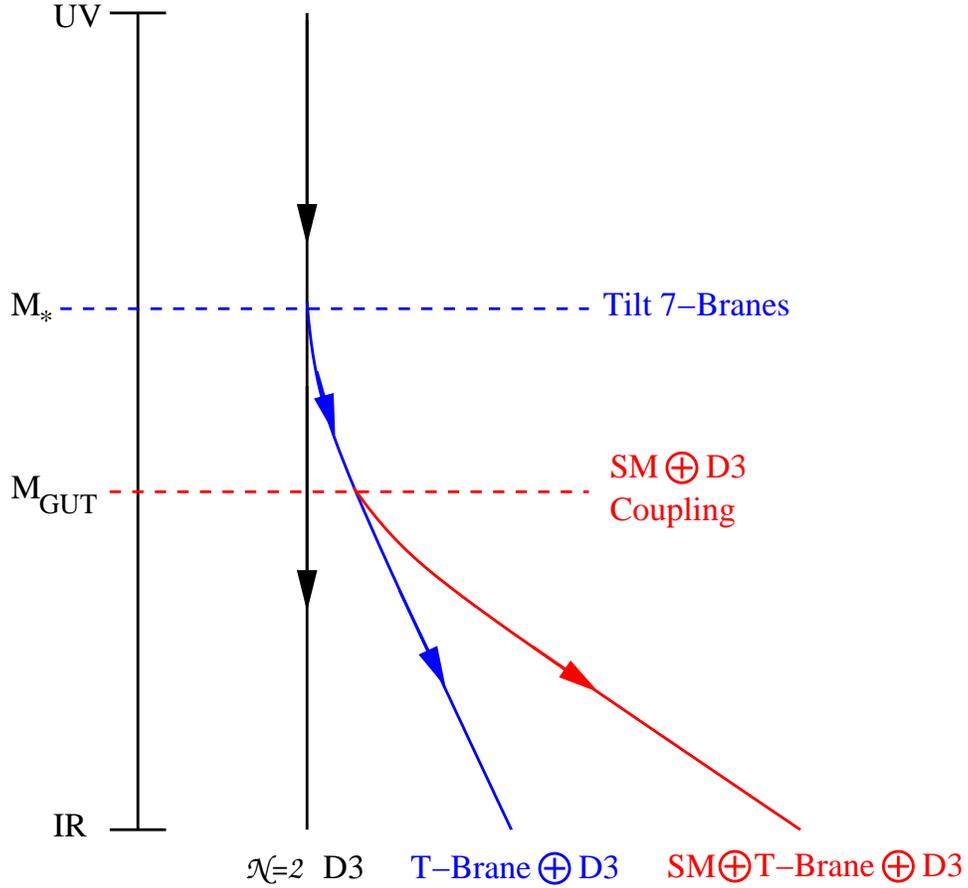}%
\caption{Depiction of the various deformations of the D3-brane probe theory.
In a limit where gravity is decoupled, in the UV, the D3-brane probes a
parallel stack of seven-branes. At a scale $M_{\ast}$, the stack of
seven-branes is tilted, inducing a flow to an interacting fixed point, denoted
by T-Brane$\oplus D3$. At a scale $M_{GUT}$, a further deformation is added,
corresponding to coupling to dynamical Standard Model fields. This leads to a
new fixed point, denoted by $SM\oplus$T-Brane$\oplus D3$ }%
\label{defflow}
\end{center}
\end{figure}

It is helpful to organize our discussion according to the energy scales of the
probe system. See figure \ref{defflow} for a depiction of the various energy scales
and corresponding deformations of the probe/MSSM system.
At high energies, when the seven-brane configuration is
parallel, the probe realizes an $\mathcal{N}=2$ theory. The effects of
seven-brane tilting are then reflected in the probe theory as a combination of
F- and D-term $\mathcal{N}=1$ deformations, which are added at the scale
$M_{\ast}$. A further deformation appears at the GUT scale, because this is
the scale at which the fields of the Standard Model become dynamical modes of
the four-dimensional theory. One of the aims of this paper is to show that in
the absence of other effects, these deformations lead to interacting fixed
points for the combined probe/MSSM system.

Throughout this work, we will have occasion to refer to the various string
states connecting the seven-branes and three-branes. We label the visible sector stack
as a $7_{SM}$-brane, and the ``flavor brane'' as a $7_{flav}$-brane. The states in the worldvolume
theory of the three-brane probe will come from the various types of $3-3$ or
$3-7$ strings. We refer to states of the probe sector which are also charged
under the SM gauge group as $3-7_{SM}$ strings, and those $3-7$ strings which
are neutral under $SU(5)_{GUT}$ as $3-7_{flav}$ strings. There are also various types
of $7-7$ strings as well, which are important for realizing the visible sector.
Let us note that this is a slight abuse of terminology, because the actual modes will involve both
weakly coupled strings, as well as $(p,q)$ strings and their junctions
\cite{Gaberdiel:1997ud, Gaberdiel:1998mv, DeWolfe:1998zf, DeWolfe:1998bi}.

The examples which have been most extensively studied in the literature
correspond to worldvolume theories which retain $\mathcal{N}=2$ supersymmetry.
In this case, there is a single stack of parallel seven-branes with gauge
group $G$. This becomes a flavor symmetry of the probe theory. A remarkable
feature of these probe theories is that the associated Seiberg-Witten curve is
given by the same F-theory geometry \cite{Sen:1996vd, Banks:1996nj}. Examples of such $\mathcal{N}=2$ probe
theories include $\mathcal{N}=2$ $SU(2)$ SYM with four flavors \cite{Banks:1996nj}, Argyres-Douglas
fixed points \cite{Argyres:1995jj}, and the Minahan-Nemeschansky theories with exceptional flavor
symmetry \cite{MNI, MNII}. Although the first example is simply a D3-brane probing a $D_4$ seven-brane, the latter two
possibilities are realized by D3-branes probing F-theory singularities of type
$H_{i}$ and $E_{n}$, and these involve intrinsically non-perturbative ingredients.
This means that such $\mathcal{N}=2$ theories are strongly coupled. The moduli
space of the $\mathcal{N}=2$ single D3-brane probe theory is given by a Higgs
branch and a Coulomb branch. These are, respectively, parameterized by the vev of a
dimension two operator $\mathcal{O}$ in the adjoint of the flavor group $G$,
and the vev of a field $Z$ which parameterizes motion normal to the seven-brane. In
addition, there is a decoupled hypermultiplet $Z_{1}\oplus Z_{2}$
parameterizing motion parallel to the seven-brane.

In the weakly coupled example given by a D3-brane probing a $D_{4}$ singularity, we
have $\mathcal{N} = 2$ super Yang-Mills theory with gauge group $SU(2)$ and
hypermultiplets $Q\oplus\tilde{Q}$. In this case, the operator
$\mathcal{O}$ is given by $\mathcal{O}\sim Q\tilde{Q}$ and has
scaling dimension two. In a theory with exceptional flavor
symmetries we lose any description of this operator in terms of elementary
fields, although one can show there are still dimension two operators
parameterizing the Higgs branch.

Tilting the seven-branes at a scale $M_{\ast}$ corresponds in the D3-brane theory
to coupling the position modes $Z_1$ and $Z_2$ to the $\mathcal{N} = 2$ theory.
This coupling breaks breaks part of the $E_8$ flavor symmetry.
Although we find the breaking $E_{8}\rightarrow SU(5)_{GUT} \times SU(5)_{\bot}$ to
be most useful for model building purposes, there is no \textit{a priori}
reason that we could not choose a different breaking pattern. This tilting
gives a mass to some of the $3-7_{flav}$ strings, and is reflected in the
probe theory as a superpotential deformation:
\begin{equation}
\delta W_{tilt}=\,\mathrm{Tr}_{G}\left(  \Phi(Z_{1},Z_{2})\cdot\mathcal{O}%
\right)  , \label{trphio}%
\end{equation}
where $\Phi(Z_{1},Z_{2})$ is valued in the adjoint of the global symmetry of
the probe theory. Choices of $\Phi$ for which $[\Phi,\Phi^{\dagger}]\neq0$
will break half of the supersymmetries, leaving $\mathcal{N}=1$ in the probe
theory. Let us note that in most realistic T-brane configurations, $[\Phi
,\Phi^{\dagger}]\neq0$ since the constant part of $\Phi$ is a non-zero
nilpotent matrix.

As found in \cite{FCFT}, superpotentials of the form (\ref{trphio}) often lead
to new strongly coupled $\mathcal{N}=1$ SCFTs in the IR. If these SCFTs
descend from $\mathcal{N}=2$ theories with exceptional global symmetries, they
will not have a known UV Lagrangian description. As emphasized in \cite{FCFT},
this does not mean that such theories are totally inaccessible to study. Since
we know the global symmetries and charges of various operators, we can still
use $a$-maximization \cite{Intriligator:2003jj} to compute the scaling dimensions of chiral primary
operators. The resulting data provides a non-trivial consistency check that such
$\mathcal{N}=1$ SCFTs exist in the IR.

Compactifying the seven-brane theory leads to an additional deformation of the
probe theory. As a first approximation, we work in the limit where the gauge
theory of the Standard Model remains as a flavor symmetry of the D3-brane
sector, but where the Standard Model fields localized on curves are dynamical.
Let us note that this is a justified approximation in F-theory GUTs, because
the gauge fields propagate over a four-dimensional worldvolume, while the
matter fields can localize on compact two-dimensional subspaces.

At the level of F-terms, the effects of compactifying the matter curves of the
system means that an additional deformation is added at the GUT scale:%
\begin{equation}
\delta W_{SM\oplus D3}=\Psi_{R}^{(SM)}\mathcal{O}_{\overline{R}}+W_{MSSM}
\label{SMdef}%
\end{equation}
for a Standard Model field $\Psi_{R}^{(SM)}$ in a representation $R$ of the
Standard Model gauge group, and operators $\mathcal{O}_{\overline{R}}$ in the
dual representation. Additionally, $W_{MSSM}$ consists of the leading order
cubic terms between the zero modes:%
\begin{equation}
W_{MSSM}=\lambda_{ij}^{(l)}H_{d}L^{i}E^{j}+\lambda_{ij}^{(d)}H_{d}Q^{i}%
D^{j}+\lambda_{ij}^{(u)}H_{u}Q^{i}U^{j}.
\end{equation}
Here, $i,j=1,2,3$ are indices running over the generations of Standard Model
fields. We neglect the contribution from the $\mu$-term and other effects
which are associated with supersymmetry breaking, as these are expected to be
induced at far lower energy scales.

The precise form of the couplings $\Psi_{R}^{(SM)}\mathcal{O}_{\overline{R}}$
depends on the details of the local profile of the matter fields, which is in
turn determined by the choice of tilting parameter $\Phi$. We review the form of these
couplings in Appendix A. See \cite{TBRANES} for further discussion.

One qualitative feature of these couplings is that in models where all three
generations localize on a single matter curve, the dominant coupling to the
probe sector will be via those modes which have maximal overlap with the
Yukawa point. Using the model of flavor physics developed in \cite{HVCKM,
BHSV, HVCP, FGUTSNC}, this maximal overlap means that the third generation (e.g. the heaviest
modes) of the Standard Model will be the ones which dominantly couple to the probe sector.
Since the Higgs up and Higgs down also naturally localize on distinct matter
curves, they will also have order one profiles at the Yukawa
point. Thus, the primary couplings are to the third generation, and the Higgs
fields. There will be some coupling to the first and second generation matter
fields, though these couplings are expected to be suppressed by powers of
$\alpha_{GUT}$. In this paper we neglect such couplings, since they are a
small correction to the basic features considered here. In this case, the
superpotential deformation can be written as:%
\begin{equation}
\delta W_{SM\oplus D3}=H_{u}\mathcal{O}_{H_{u}}+H_{d}\mathcal{O}_{H_{d}%
}+\overline{5}_{M}\mathcal{O}_{\overline{5}_{M}}+10_{M}\mathcal{O}_{10_{M}%
}+W_{MSSM}%
\end{equation}
in the obvious notation.

Gauge field interactions constitute another source of interaction between the
probe and the Standard Model. For example, since the probe contains states
charged under the Standard Model gauge group, these modes will couple to the
corresponding gauge fields, affecting the running of couplings. Additionally,
kinetic mixing between the probe and Standard Model is generically expected
\cite{Funparticles}. All of these couplings lead to a very rich quasi-hidden
sector. One of our aims in this paper will be to quantify some details of the mixed
probe/MSSM system.

\section{Fixed Points of the Coupled Probe/MSSM System\label{sec:FIXED}}

In the previous section we reviewed the basic setup for the system described
by a strongly coupled D3-brane probing the Standard Model. In this section we
consider a particular idealization where we do not include effects from breaking
either the conformal symmetry or supersymmetry. Further, we work in the limit where the
SM gauge group can be treated as a flavor symmetry so that it is
only weakly gauged. However, we do allow the Standard Model fields lcoalized on
curves to be dynamical. Geometrically, this corresponds to a limit where some
of the matter curves are compact, while the full GUT\ seven-brane is non-compact.

Under these circumstances, we provide evidence that the coupled probe/MSSM
systems will flow to non-trivial $\mathcal{N}=1$ superconformal
fixed points. The precise fixed point in question is controlled by the choice
of UV deformation, which we summarize as:%
\begin{equation}
\delta W_{UV}=\delta W_{tilt}+\delta W_{SM\oplus D3}\text{.}%
\end{equation}
This depends on the choice of $\Phi$, as well as the zero mode content on the
matter curves. Perhaps surprisingly, we find that not only does the
coupled probe/MSSM system admit such fixed points, but that the operators of
the Standard Model develop only \textit{small} anomalous dimensions.
Due to the small shift in the scaling dimensions of Standard Model fields, there is a well-defined sense
in which the Standard Model degrees of freedom retain their identity. This is
important, because it means that once we include the effects of CFT breaking,
the identity of the Standard Model states remains roughly the same as in the
UV description in terms of intersecting seven-branes. The low amount of mixing is
further corroborated by the computable effect of the probe on the running of the SM gauge couplings.
Indeed, we find only a mild (but irrational) shift to the usual one-loop beta functions.

In much of this paper, we shall focus on the effects of a single probe D3-brane, though we shall
also consider the case of two D3-branes as well. Indeed, adding a large number of D3-branes
induces a much larger threshold correction to the running of the SM gauge couplings. This in
turn can induce a Landau pole before the GUT scale. In the context of F-theory GUTs which
admit a decoupling limit \cite{BHVII, HVGMSB, HeckVerlinde}, there is actually a more stringent requirement
that the zero mode content must preserve asymptotic freedom of $SU(5)_{GUT}$.
This limits the amount of additional matter which can be added below the GUT scale.
To get a rough sense of the amount of allowed extra matter at low energies,
we can consider the beta function associated with three generations of
chiral matter, one vector-like pair of $5_{H} \oplus \overline{5}_{H}$, and $\delta b_{SU(5)}$
additional GUT multiplets in the $5 \oplus \overline{5}$. The resulting beta function
is, in our sign conventions:
\begin{equation}
b_{SU(5)} = - 15 + 7 + \delta b_{SU(5)}.
\end{equation}
In other words, if we demand $b_{SU(5)} < 0$, we obtain the condition
$\delta b_{SU(5)} < 8$. In practice, we find that in most probe scenarios, this bound
limits us to one or two probe D3-branes.

The rest of this section is organized as follows. By appealing to the symmetries of our
non-Lagrangian probe theory we determine up to one unfixed parameter the
general form of the infrared R-symmetry. Much as in \cite{FCFT}, this parameter
is then fixed by $a$-maximization \cite{Intriligator:2003jj}. Determining the infrared R-symmetry
allows us to compute the scaling dimensions of chiral primary
operators such as the MSSM chiral superfields. We also show that
in the limit where the Standard Model gauge group is treated as a weakly gauged flavor symmetry,
we can also extract the change in the beta functions from the probe theory. After this, we present in
detail a particular example of a probe theory. This is followed by a summary of various probe scenarios involving
one, as well as two probe D3-branes.

\subsection{Infrared R-Symmetry}

\bigskip In general, knowing the global symmetries of an SCFT is enough to
figure out the R-charges (or equivalently the scaling dimensions) of all
chiral primary operators. The procedure for doing this was first described in
\cite{Intriligator:2003jj}, in which it was discovered that the superconformal
R-symmetry is the one that locally maximizes
\begin{equation}
a_{trial}=\frac{3}{32}\left(  3\,\mathrm{Tr}R_{trial}^{3}-\,\mathrm{Tr}%
R_{trial}\right)  , \label{atrial}%
\end{equation}
where $R_{trial}=R_{0}+\sum_{I}s_{I}F_{I}$ is a general linear combination of
an R-symmetry $R_{0}$ and all anomaly-free global symmetries $F_{I}$. Finding
the unique local maximum of (\ref{atrial}) then fixes all the coefficients
$s_{I}$.

Assuming the absence of accidental IR\ symmetries, we know the set of global
symmetries of the SCFTs we study in this work. These are given by the
symmetries of the original $\mathcal{N}=2$ probe theory, and additional
symmetries which act on the Standard Model. The various charges of the probe
sector and Standard Model fields, in the case where we begin with an $E_8$ $\mathcal{N}=2$ theory, are summarized below%
\begin{equation}
\label{tabtab}%
\begin{tabular}
[c]{|c|c|c|c|c|c|}\hline
UV symmetries & $\mathcal{O}_{s}$ & $Z$ & $Z_{1}$ & $Z_{2}$ & $\Psi_{SM}%
$\\\hline
$R_{UV}$ & $\frac{4}{3}$ & $4$ & $\frac{2}{3}$ & $\frac{2}{3}$ & $\frac{2}{3}%
$\\\hline
$J_{\mathcal{N}=2}$ & $-2$ & $12$ & $-1$ & $-1$ & $0$\\\hline
$U(1)_{1}$ & $0$ & $0$ & $+1$ & $0$ & $0$\\\hline
$U(1)_{2}$ & $0$ & $0$ & $0$ & $+1$ & $0$\\\hline
$U(1)_{\Psi}$ & $0$ & $0$ & $0$ & $0$ & $+1$\\\hline
$T_{3}$ & $s$ & $0$ & $0$ & $0$ & $s_{\Psi}$\\\hline
\end{tabular}
\ .
\end{equation}
Here, $R_{UV}$ denotes the UV R-symmetry of the theory when treated in terms
of a decoupled $\mathcal{N}=2$ system and a weakly coupled set of free fields
for the Standard Model degrees of freedom. In the context of an $\mathcal{N}%
=2$ superconformal field theory, $R_{UV}$ and $J_{\mathcal{N}=2}$
combine to form an abelian subalgebra of the $SU(2)\times U(1)$ R-symmetry. The unusual $J_{\mathcal{N}=2}$ charge of $Z$ is a result of using the $E_8$ theory as our starting point. We
have also indicated the rephasing symmetries for the decoupled hypermultiplet
$Z_{1}\oplus Z_{2}$. On similar grounds, we have also included a set of
rephasing symmetries for the various Standard Model fields. For each Standard
Model field, there is a corresponding $U(1)_{\Psi}$. We note that although
such symmetries will be broken by the presence of cubic interaction terms in
$W_{MSSM}$, such interaction terms do not affect the IR fixed point since they
are irrelevant in the IR.\footnote{In more
precise terms, we note that the MSSM\ superpotential consists of terms cubic
in Standard\ Model operators, since the $\mu$-term is assumed to be generated
at scales below the CFT\ breaking scale. Given a cubic coupling, the only way
that this term can be maintained as a marginal coupling in the IR is if
all Standard Model fields in the interaction term remain free fields. If this
occurs, however, then these Standard Model modes are effectively decoupled
from the probe anyway.}

Finally, there is also a symmetry generator $T_{3}$ associated with the
tilting of the seven-brane configuration. This generator is given as a linear
combination of generators in the Cartan subalgebra of the parent group
$G_{parent}$. The precise definition is as follows \cite{FCFT}. Given a nilpotent Jordan
block decomposition of the constant part of $\Phi$:%
\begin{equation}
\Phi_{0}=\underset{a=1}{\overset{k}{\oplus}}J^{(a)},
\end{equation}
each block has an associated spin $j_{(a)}=\frac{1}{2}(n_{(a)}-1)$
representation of $SU(2)$. Denote by $T_{3}^{(a)}$ the $L_{z}$ generator of
this representation. The generator $T_{3}$ is then given as a sum of these
generators:%
\begin{equation}
T_{3}\equiv\sum_{a}T_{3}^{(a)}.
\end{equation}
In line (\ref{tabtab}) we have denoted the $T_{3}$ charge of an operator
$\mathcal{O}_{s}$ by $+s$, and that of a Standard Model field $s_{\Psi}$. Let
us note that here, we are working with respect to a particular holomorphic gauge in which
the SM field has a well-defined $T_{3}$ charge \cite{TBRANES}. See Appendices A and B for further
discussion on this point.

In terms of these symmetries, the infrared R-symmetry can be determined, much
as in \cite{FCFT}, to be a linear combination of the form:%
\begin{equation}
R_{IR}=R_{UV}+\left(  \frac{t}{2}-\frac{1}{3}\right)  J_{\mathcal{N}=2}%
-tT_{3}+u_{1}U_{1}+u_{2}U_{2}+\underset{\Psi}{\sum
}u_{\Psi}U_{\Psi}.
\end{equation}
Here the sum over $\Psi$ extends over all MSSM chiral superfields which
couple to the probe brane sector.

The $u_{i}$ coefficients are fixed by the condition that some of the
deformations in $\delta W_{tilt}$ are marginal in the infrared. For example, the
coefficients $u_{1}$ and $u_{2}$ are:
\begin{equation}
u_{1}  =\left(  S_{1}+\frac{3}{2}\right)  t-1 \,\,,\,\, u_{2} =\left(  S_{2}+\frac{3}{2}\right)  t-1
\end{equation}
where $S_{1}$ (resp. $S_{2}$) denotes the contribution from the
deformation $\Phi(Z_{1},Z_{2})$ which is linear in $Z_{1}$ (resp. $Z_{2}$),
and has the lowest $T_{3}$ charge. We refer to the coefficients appearing in the above as:
\begin{align}
\mu_{1}  =\left(  S_{1}+\frac{3}{2}\right) \,\,,\,\, \mu_{2}  =\left(  S_{2}+\frac{3}{2}\right).
\end{align}

Similar considerations allow us to fix the $u_{\Psi}$ coefficients.
The basic condition is that if we demand that the operator
$\Psi_{R}\cdot\mathcal{O}_{\overline{R}}$ is marginal in the infrared, then we
must require $R_{IR}(\Psi_{R}^{SM}\cdot\mathcal{O}_{\overline{R}})=2$. In
holomorphic gauge, $\Psi_{R}$ and $\mathcal{O}_{\overline{R}}$ have opposite
$T_{3}$ charges \cite{TBRANES}. This imposes the condition:%
\begin{equation}\label{uPSI}
u_{\Psi}=t-\frac{2}{3}.
\end{equation}
To summarize, we introduce a net $U(1)_{SM}$ given by the sum of all
$U(1)_{\Psi}$'s which are not free fields in the infrared:%
\begin{equation}
U\left(  1\right)  _{SM}=\underset{\Psi\text{ not free}}%
{\sum}U(1)_{\Psi}.
\end{equation}
The infrared R-symmetry is then:%
\begin{equation}
R_{IR}=R_{UV}+\left(  \frac{t}{2}-\frac{1}{3}\right)  J_{\mathcal{N}=2}%
-tT_{3}+u_{1}U_{1}+u_{2}U_{2}+\left(  t-\frac{2}{3}\right)  U_{SM}.
\label{ircharges}
\end{equation}
Thus, the infrared R-symmetry is determined up to a single parameter $t$ which
is fixed by $a$-maximization.

To perform $a$-maximization, we need to evaluate the cubic and
linear anomalies in $R_{IR}$. The computation is quite similar to the one
presented in \cite{FCFT}. The main idea is that although it is difficult to
compute the cubic and linear anomalies $R_{IR}^{3}$ in the IR, since
these symmetry currents can be seen in the UV, we can via anomaly matching
determine the form of these expressions in the UV as well. In the UV, however,
note that the degrees of freedom of the Standard\ Model and the probe sector
have decoupled. Hence, the cubic and linear anomalies are given by:%
\begin{align}
R_{IR}^{3}  &  =\Tr_{D3}R_{IR}^{3}+\Tr_{SM}R_{IR}^{3}.\\
R_{IR}  &  =\Tr_{D3}R_{IR}+\Tr_{SM}R_{IR}.
\end{align}
Here, $\Tr_{D3}$ refers to evaluating the corresponding anomaly by tracing over
just the degrees of freedom given by the UV $\mathcal{N}=2$ D3-brane probe
theory. The second contribution refers to the trace over those Standard Model
degrees of freedom which mix non-trivially with the probe in the infrared. The
key point is that because these two systems are decoupled in the UV, we can
evaluate these contributions separately.

Consider first the trace over the UV D3-brane degrees of freedom. In this
UV\ theory, there are no states charged under $U(1)_{SM}$. In other words, the
evaluation of the cubic anomaly as a function of $t$ is identical to that
already given in \cite{FCFT}. The resulting expressions are:%
\begin{align}
\Tr_{D3}R_{IR}^{3}  &  =\left(  12a_{E_{8}}-9c_{E_{8}}-\frac{3k_{E_{8}}r}{4}\right)
t^{3}+\left(  -24a_{E_{8}}+12c_{E_{8}}+\frac{3u_{1}}{4}+\frac{3u_{2}}{4}\right)
t^{2}\\
&  +\left(  24a_{E_{8}}-12c_{E_{8}}-\frac{3u_{1}^{2}}{2}-\frac{3u_{2}^{2}}%
{2}\right)  t+(u_{1}^{3}+u_{2}^{3})\\
\Tr_{D3}R_{IR}  &  =(24a_{E_{8}}-24c_{E_{8}})t+(u_{1}+u_{2}).
\end{align}
Here, $r$ is a group theory parameter which measures the size of the Jordan block structure associated with $\Phi(0,0)$:
\begin{equation}
r = 2 \Tr(T_{3}T_{3}).
\end{equation}
The central charges $a_{E_{8}}$, $c_{E_{8}}$ and $k_{E_{8}}$ are the anomaly
coefficients associated with the $\mathcal{N}=2$ SCFT in the UV. For a probe
with $N$ D3-branes, the resulting values are
\cite{Cheung:1997id,Aharony:2007dj}:
\begin{align}
a_{E_{n}}  &  =\frac{1}{4}N^{2}\Delta+\frac{1}{2}N(\Delta-1),\\
c_{E_{n}}  &  =\frac{1}{4}N^{2}\Delta+\frac{3}{4}N(\Delta-1),\\
k_{E_{n}}  &  =2N\Delta.
\end{align}
where $N$ is the number of coincident probe D3-branes and $\Delta=6,4,3$ is
the scaling dimension of the Coulomb branch parameter for the $E_{8}$, $E_{7}$, and
$E_{6}$ Minahan-Nemeschansky theories, respectively. Here we have included the
contribution from the decoupled hypermultiplet $Z_1 \oplus Z_2$, which is why our expression is
different from the one in \cite{Aharony:2007dj}.  In some cases such as where
only $Z_{1}$ couples to the configuration, there is an additional factor which
must be subtracted.

The contribution from the MSSM degrees of freedom can also be evaluated in
the UV:
\begin{align}
\Tr_{SM}R_{IR}^{3}  &  =\underset{\Psi\text{ not free}}{\sum}d_{\Psi}%
\times\left(  -\frac{1}{3}-t\cdot T_{3}(\Psi)+u_{\Psi}\right)  ^{3}\\
\Tr_{SM}R_{IR}  &  =\underset{\Psi\text{ not free}}{\sum}d_{\Psi}\times\left(
-\frac{1}{3}-t\cdot T_{3}(\Psi)+u_{\Psi}\right)
\end{align}
where $d_{\Psi}$ is the dimension of the representation for $\Psi$.
Putting this together, we can write the trial central charge $a_{trial}$ as:%
\begin{equation}
a_{trial}=\frac{3}{32}\left(  3\Tr_{D3}R_{IR}^{3}+3\Tr_{SM}R_{IR}^{3}%
-\Tr_{D3}R_{IR}-\Tr_{SM}R_{IR}\right)  .
\end{equation}
$a$-maximization then implies that the local maximum of $a_{trial}$ as a
function of $t$ yields the value of $t=t_{\ast}$ corresponding to the infrared superconformal R-symmetry.

One consequence of this analysis is that we typically find that the Standard
Model fields develop only small anomalous dimensions. This is important because
it means that the Standard Model fields basically retain
their identity, even in the infrared theory. As we show later, this also means that in
more realistic situations, the actual suppression of the MSSM superpotential
will not be that significant. Assuming flavor is generated at the GUT scale, this
allows us to lower the CFT breaking scale below the GUT scale.\footnote{Let us note that it is in principle possible to
consider models of flavor physics where hierarchical mass patterns are generated by non-zero anomalous
dimensions \cite{Nelson:2000sn,Nelson:2001mq}. This is also a logical possibility in the present class of models. Here,
the idea would be that what is referred to as a ``third generation field'' at the GUT scale develops conformally suppressed Yukawas. Since what is
referred as the ``second generation field'' at the GUT scale couples only weakly to the probe D3-brane, this mode would, at lower energies,
become the effective third generation. Note that this leads to some suppression in the overall Yukawa matrices, unless additional fine-tuning
is included. Though it would be interesting to study such possibilities further, in this work we mainly consider the most straightforward option
that what is identified as the third generation at the GUT scale remains so at lower energies as well.}

As a final remark, let us note that there is a further class of deformations one can consider adding to the CFT,
given by allowing some of the Standard Model fields to develop non-zero vevs. For example, a non-zero Higgs vev
can trigger CFT breaking for the D3-brane sector. In this case, one can see that the resulting theory does not flow
to an interacting fixed point. This is analogous to what happens in SQCD when enough flavors get a mass to push
the theory out of the conformal window. Indeed, returning to our discussion
of the infrared R-symmetry in equation (\ref{ircharges}), we see that if we demand the deformation
$\langle \Psi_{R} \rangle \cdot \mathcal{O}_{\overline{R}}$ has R-charge $+2$ in the infrared, then the
parameter $t$ satisfies the condition $t(s + 1) = 0$ where $s$ is the $T_{3}$ charge of $\mathcal{O}_{\overline{R}}$. On
the other hand, in all T-brane examples, localized matter fields have $s < 0$, so the $\mathcal{O}$'s that they can pair
with have $s > 0$. This means that $t = 0$, which is clearly problematic for an interacting fixed point. This is
an indication that the theory is no longer conformal and instead develops a characteristic mass scale on the
order of $\langle \Psi_{R} \rangle$.

\subsection{Beta Functions}

Assuming that we do flow to an interacting CFT, we would like to establish
certain properties about how it affects the visible sector. To this end, we
now discuss how the probe sector affects the running of the visible sector
gauge couplings.\ Under the assumption (soon to be verified in a variety of
examples) that the Standard Model is only perturbed slightly by coupling to
the CFT, we can compute at weak gauge coupling the effects of an additional
threshold correction from the new states charged under the Standard
Model gauge group. From the perspective of the CFT, this amounts to computing
the two-point function for the flavor symmetry currents of $SU(3)\times
SU(2)\times U(1)$. Let us denote these currents by $J_{G}$ for $G=SU(3)$,
$SU(2)$, $U(1)$. Upon weakly gauging this current,
the one loop beta function will then be given by :%
\begin{equation}
\beta_{G}\equiv\frac{\partial g_{G}}{\partial\ln\mu}=\frac{g_{G}^{3}}%
{16\pi^{2}}b_{G},\quad\mathrm{where}\quad b_{G}=-3\,\mathrm{Tr}(R_{IR}%
J_{G}J_{G}). \label{betafunc}%
\end{equation}
Here, the \textquotedblleft Tr\textquotedblright\ trace refers to the anomaly
coefficient associated with one $R_{IR}$ current and two $J_{G}$ global
symmetry currents. In a weakly coupled setting, the trace would be over the
elementary degrees of freedom of the theory.\ This formula comes from treating
the CFT as matter interacting with the weakly gauged flavor
symmetry\footnote{For a detailed discussion, see \cite{Benini:2009mz}.}.

As a first warmup case, we can consider the contribution to the running of the $SU(5)_{GUT}$ coupling
in the limit where there are no probe/MSSM F-term couplings. This corresponds to the case where only $\delta W_{tilt}$ enters
as a deformation of the probe theory. This limiting case has been studied in \cite{FCFT}. Letting $t_{\Phi}$ denote the value of
the parameter $t$ obtained by performing $a$-maximization in this simplified case, the universal GUT contribution
from the probe D3-brane is:
\begin{equation}\label{deltabsufiv}
\delta b_{SU(5)} \equiv \frac{3 k_{E_{8}} t_{\Phi}}{4}.
\end{equation}
We shall encounter a similar contribution later for each Standard Model gauge group factor. However, to emphasize the fact that this is
associated with a GUT scale threshold, we reserve the notation $\delta b_{SU(5)}$ for equation (\ref{deltabsufiv}).

We now compute the beta function for the Standard Model gauge group
factors. Near the scale where the MSSM superpartners first enter the
spectrum, the corresponding one-loop MSSM\ beta functions are, in our sign conventions:%
\begin{equation}
b_{SU(3)}^{(0)}=-3\text{, }b_{SU(2)}^{(0)}=+1\text{, }b_{U(1)}^{(0)}%
=+\frac{33}{5}.
\label{origbeta}
\end{equation}
In a step function approximation to the running, this is the value of the beta
function until the additional states from the probe sector enter the spectrum.
At this point, additional contributions will affect the running.

Turning the discussion around, we can proceed from high energies down to
lower energies. There will then be the usual contribution from $b^{(0)}$, as
well as additional contributions induced by the presence of the probe sector.
As indicated in equation (\ref{betafunc}), the one-loop beta function is given
by an anomaly coefficient. In the IR, it is difficult to directly list all of
the degrees of freedom, because the probe and Standard Model now interact
non-trivially. Note, however, that because $b_{G}$ is an anomaly coefficient,
we can use anomaly matching to compute it in the UV, where the Standard Model
and probe sector degrees of freedom are decoupled. The key point for us is
that $R_{IR}$ is a linear combination of $R_{UV}$ and flavor symmetries of the
UV theory, and $J_{G}$ is a flavor symmetry present in both the UV\ and IR
theories.

In terms of the UV degrees of freedom, we then have:%
\begin{equation}
b_{G}=-3\,\mathrm{Tr}(R_{IR}J_{G}J_{G})=-3\,\mathrm{Tr}_{SM}(R_{IR}J_{G}%
J_{G})-3\,\mathrm{Tr}_{D3}(R_{IR}J_{G}J_{G}).
\end{equation}
Using the form of the IR R-symmetry (\ref{ircharges}) and the charge assignments in (\ref{tabtab}),
we find that the shifts in the beta functions
\begin{equation}
\delta b_{G} \equiv b_{G}-b_{G}^{(0)}%
\end{equation}
for a gauge group factor $G$ are given by:%
\begin{align}
\delta b_{SU(3)}  &  =\frac{3k_{E_{8}}}{4}t+\frac{9}{2}
\times\left(  t\cdot T_{3}^{(10_{M})}-u_{10_{M}}\right) +\frac{3}{2}\times\left(  t\cdot
T_{3}^{(\overline{5}_{M})}-u_{\overline{5}_{M}}\right)  \\
\delta b_{SU(2)}  &  =\frac{3k_{E_{8}}}{4}t+\frac{9}{2}%
\times\left(  t\cdot T_{3}^{(10_{M})}-u_{10_{M}}\right) +\frac{3}{2}\times\left(  t\cdot
T_{3}^{(\overline{5}_{M})}-u_{\overline{5}_{M}}\right)  \\
&  \qquad\qquad\,\,+\frac{3}{2}\times\left(  t\cdot T_{3}^{(H_{d})}-u_{H_{d}%
}\right)  +\frac{3}{2}\times\left(  t\cdot T_{3}^{(H_{u})}-u_{H_{u}}\right) \\
\delta b_{U(1)}  &  =\frac{3k_{E_{8}}}{4}t+\frac{9}{2}%
\times\left(  t\cdot T_{3}^{(10_{M})}-u_{10_{M}}\right) +\frac{3}{2}\times\left(  t\cdot
T_{3}^{(\overline{5}_{M})}-u_{\overline{5}_{M}}\right)  \\
&  +3\times\frac{18}{60}\times\left(  t\cdot T_{3}^{(H_{d})}-u_{H_{d}}\right)
+3\times\frac{18}{60}\times\left(  t\cdot T_{3}^{(H_{u})}-u_{H_{u}}\right)  .
\end{align}
Here, we are assuming that all modes have developed an anomalous dimension. The corresponding terms which are
free fields are to be omitted from this expression. In all cases, the first line is the same for $SU(3)$, $SU(2)$ and $U(1)$. In particular, such contributions will not distort gauge coupling unification. The additional contributions from the Higgs fields to each gauge group factor are
different. We shall later comment on the effects of these shifts, and their (helpful)
consequences for precision unification. Note also that these shifts are missing the contribution
from $R_{UV}$ in (\ref{ircharges}), since that effect is contained in $b_G^{(0)}$ and has
been subtracted off. Finally, note that our normalization for the $U(1)$ generator is chosen to agree with a canonical
embedding in $SU(5)_{GUT}$.

\subsection{Example: $\mathbb{Z}_{2}$ Monodromy}

To illustrate the main ideas discussed earlier, in this section
we consider a simplified model based on a $\mathbb{Z}_{2}$
monodromy group which realizes a visible sector with the couplings
$5_H \times 10_M \times 10_M$ and $\overline{5}_H \times \overline{5}_M \times 10_M$. These
examples do not include a neutrino scenario as in \cite{BHSV,EPOINT}. However,
given the extra flexibility afforded by D3-branes, there may be novel ways to
include neutrinos in such setups. Since they are among the simplest examples
of T-brane configurations we consider these examples first.

The coarse-grained T-brane configuration we consider is defined by:
\begin{equation}
\Phi=\left[
\begin{array}
[c]{ccccc}%
0 & 1 &  &  & \\
Z_{1} & 0 &  &  & \\
&  & 0 &  & \\
&  &  & 0 & \\
&  &  &  & 0
\end{array}
\right]  . \label{deformar}%
\end{equation}
The characteristic polynomial for $\Phi$ has Galois group $\mathbb{Z}_{2}$, which is identified with the ``monodromy group''.
Higher order (irrelevant to the IR D3-brane theory) terms serve to fix the profile of
localized matter in the geometry. As explained in Appendix A, such
considerations are not so important for the considerations of this paper. The
$T_{3}$ generator is given by:%
\begin{equation}
T_{3}=\text{diag}(1/2,-1/2,0,0,0).
\end{equation}

In a singular branched gauge, we can describe this scenario as having a $%
\mathbb{Z}
_{2}$ monodromy group which acts on the eigenvalues $\lambda_{i}$ of $\Phi$
by permuting $\lambda_{1}$ and $\lambda_{2}$, while keeping the other $\lambda_{i}$ fixed.
In terms of the eigenvalues $\lambda_{i}$,
we have that the visible sector modes fill out the
following orbits under the $%
\mathbb{Z}
_{2}$ monodromy group:%
\begin{align}
10_{M}  &  :\{\lambda_{1},\lambda_{2}\}\\
5_{H}  &  :\{-\lambda_{1}-\lambda_{2}\}\\
\overline{5}_{1}  &  :\{\lambda_{1}+\lambda_{3},\lambda_{2}+\lambda_{3}\}\\
\overline{5}_{2}  &  :\{\lambda_{4}+\lambda_{5}\}.
\end{align}
The subscript on the $\overline{5}$ fields indicates that in this simple
example, we can interchange the roles of the $\overline{5}_{M}$ and
$\overline{5}_{H}$. We consider both possibilities in what follows, and refer to the monodromy
scenarios as $\mathbb{Z}^{(1)}_{2}$ and $\mathbb{Z}^{(2)}_{2}$.

Based on our general discussion of probe/MSSM couplings in Appendix A, we can
determine the vector transforming in a representation of $SU(5)_{\bot}$ which
dominantly couples to the D3-brane, as well as its corresponding $T_{3}$
charge:
\begin{align}
&
\begin{tabular}
[c]{|c|c|c|c|c|}\hline
$%
\mathbb{Z}
_{2}^{(1)}$ & $H_{u}$ & $H_{d}$ & $\overline{5}_{M}$ & $10_{M}$\\\hline
Vector & $e_{1}^{\ast}\wedge e_{2}^{\ast}$ & $e_{2}\wedge e_{3}$ &
$e_{4}\wedge e_{5}$ & $e_{2}$\\\hline
$T_{3}$ charge & $0$ & $-1/2$ & $0$ & $-1/2$\\\hline
\end{tabular}
\\
&
\begin{tabular}
[c]{|c|c|c|c|c|}\hline
$%
\mathbb{Z}
_{2}^{(2)}$ & $H_{u}$ & $H_{d}$ & $\overline{5}_{M}$ & $10_{M}$\\\hline
Vector & $e_{1}^{\ast}\wedge e_{2}^{\ast}$ & $e_{4}\wedge e_{5}$ &
$e_{2}\wedge e_{3}$ & $e_{2}$\\\hline
$T_{3}$ charge & $0$ & $0$ & $-1/2$ & $-1/2$\\\hline
\end{tabular}
\text{ \ \ \ }.
\end{align}

\subsubsection*{$\Phi$-deformed Theory}

To begin our analysis of the probe/MSSM fixed points, we first recall some
properties of the CFT obtained in the limit in which all Standard Model fields
are decoupled. In other words, we first study the theory obtained by the
deformation%
\begin{equation}
\delta W_{tilt} =\Tr_{E_{8}}\left(  \Phi(Z_{1},Z_{2})\cdot\mathcal{O}\right)
\end{equation}
with $\Phi$ given as in equation (\ref{deformar}). The resulting infrared
fixed point has basically been obtained in \cite{FCFT}. The parameters
$\mu_{1}=5/2$, $r=1$. The value of $\mu_{2}$ has \ no physical meaning in this
case, because $Z_{2}$ remains decoupled in the infrared. Via $a$-maximization,
we can determine the critical value of $t$ which maximizes $a$, which we
denote by $t_{\Phi}$. Its value is $t_{\Phi} = 0.53$.

We can also compute the infrared values of the central
charges $a$, $c$ and $k$. To properly compare different theories, we also
include the contributions to $a$ and $c$ from the UV\ decoupled modes $Z_{2},$
as well as the decoupled Standard Model fields $H_{u}$, $H_{d}$, $\overline
{5}_{M}$ and $10_{M}$. Finally, we also consider the contribution to the
running of the $SU(5)_{GUT}$ coupling from just the D3-brane sector, given by
$\delta b_{SU(5)}=3k_{E_{8}}t_{\ast}/4$:%
\begin{equation}%
\begin{tabular}
[c]{|c|c|c|c|c|}\hline
& $t_{\Phi}$ & $a_{\Phi}$ & $c_{\Phi}$ & $\delta b_{SU(5)}$\\\hline
$%
\mathbb{Z}
_{2}$ & $0.53$ & $3.83$ & $5.21$ & $4.80$\\\hline
\end{tabular}
\end{equation}

Given the values of these parameters, we can also determine the scaling
dimensions of the operators which eventually will couple to the Standard Model
fields. This is dictated by the $T_{3}$ charge assignments, and so will depend
on the particular monodromy scenario we consider. In our case, we have
operators with $T_{3}$ charge $0$ or $+1/2$. This leads to scaling dimensions:%
\begin{align}
\Delta\left(  \mathcal{O}_{s=0}\right)   &  =3-\frac{3}{2}t\\
\Delta\left(  \mathcal{O}_{s=1/2}\right)   &  =3-\frac{9}{4}t
\end{align}

We can use this to determine which of our original deformations are relevant,
or marginal in the IR\ of the $\Phi$-deformed theory:
\begin{equation}%
\begin{tabular}
[c]{|c|c|c|c|c|}\hline
Dimensions & $\mathcal{O}_{H_{u}}$ & $\mathcal{O}_{H_{d}}$ & $\mathcal{O}%
_{\overline{5}_{M}}$ & $\mathcal{O}_{10_{M}}$\\\hline
$%
\mathbb{Z}
_{2}^{(1)}$ & $2.20$ & $1.80$ & $2.20$ & $1.80$\\\hline
$%
\mathbb{Z}
_{2}^{(2)}$ & $2.20$ & $2.20$ & $1.80$ & $1.80$\\\hline
\end{tabular}
\ \ \ \ \ \ \ .
\end{equation}

\subsubsection*{Infrared Theory}

Let us now compute the form of the infrared R-symmetry in the two $\mathbb{Z}_{2}$
monodromy scenarios. Again, we must evaluate the various cubic and linear anomalies.
Here, we have:
\begin{align}
\mathbb{Z}^{(1)}_{2}\text{ Case}  &  \text{: }R_{IR}^{3}=\Tr_{D3}R_{IR}%
^{3}+(2+10)\times\left(  -\frac{1}{3}+\frac{t}{2}+u_{\Psi}\right)  ^{3}\\
\mathbb{Z}_{2}^{(2)}\text{ Case}  &  \text{: }R_{IR}^{3}=\Tr_{D3}R_{IR}%
^{3}+(5+10)\times\left(  -\frac{1}{3}+\frac{t}{2}+u_{\Psi}\right)  ^{3}%
\end{align}
where in the above $u_{\Psi} = t - 2/3$ as in equation (\ref{uPSI}).
Similarly, we can evaluate the anomaly linear in R-charge. We have:%
\begin{align}
\mathbb{Z}_{2}^{(1)}\text{ Case}  &  \text{: }R_{IR}=\Tr_{D3}R_{IR}%
+(2+10)\times\left(  -\frac{1}{3}+\frac{t}{2}+u_{\Psi}\right) \\
\mathbb{Z}_{2}^{(2)}\text{ Case}  &  \text{: }R_{IR}=\Tr_{D3}R_{IR}%
+(5+10)\times\left(  -\frac{1}{3}+\frac{t}{2}+u_{\Psi}\right)  .
\end{align}
Performing $a$-maximization in these two cases yields the same value (after
rounding):%
\begin{equation}
t_{\ast}=0.50. \label{tmaxone}%
\end{equation}
From this, we conclude that the IR\ dimensions of the Standard Model fields,
and $\mathcal{O}$ operators are:%
\begin{equation}%
\begin{tabular}
[c]{|c|c|c|c|c|c|c|c|c|}\hline
IR\ Dimensions & $H_{u}$ & $\mathcal{O}_{H_{u}}$ & $H_{d}$ & $\mathcal{O}%
_{H_{d}}$ & $\overline{5}_{M}$ & $\mathcal{O}_{\overline{5}_{M}}$ & $10_{M}$ &
$\mathcal{O}_{10_{M}}$\\\hline
$%
\mathbb{Z}
_{2}^{(1)}$ & $1$ & $2.25$ & $1.13$ & $1.87$ & $1$ & $2.25$ & $1.13$ &
$1.87$\\\hline
$%
\mathbb{Z}
_{2}^{(2)}$ & $1$ & $2.25$ & $1$ & $2.25$ & $1.12$ & $1.88$ & $1.12$ &
$1.88$\\\hline
\end{tabular}
\ \ \ \ \ \ \ \ \ .
\end{equation}
Let us consider the suppression of the Yukawa couplings associated
with the superpotential couplings~$W_{MSSM}$. For the modes which have maximal
coupling to the probe D3-brane sector, we have, at the CFT breaking scale:%
\begin{equation}
\lambda_{5\times10\times10}\sim\lambda_{\overline{5}\times\overline{5}%
\times10}\sim\left(  \frac{M_{\cancel{CFT}}}{M_{GUT}}\right)  ^{0.26}.
\end{equation}
For $M_{\cancel{CFT}}/M_{GUT}\sim10^{-3}$, this leads to $\lambda\sim0.2$,
which is only a mild suppression. This is easily compensated for
by a small enhancement of the Yukawa at the GUT scale.

We can also compute the values of the central charges $a$ and $c$, and compare
with their values in the original $\mathcal{N}=2$ and $\Phi$-deformed
theories. Here, the UV theory contains both the contributions to $a$ and $c$
from the $\mathcal{N}=2$ probe D3-brane, as well as the decoupled
Standard\ Model fields and the decoupled hypermultiplet $Z_{1}\oplus Z_{2}$.
We find:%
\begin{equation}%
\begin{tabular}
[c]{|c|c|c|c|c|c|c|}\hline
Central Charges & $a_{UV}$ & $a_{\Phi}$ & $a_{IR}$ & $c_{UV}$ & $c_{\Phi}$ &
$c_{IR}$\\\hline
$\mathbb{Z}_{2}^{(1)}$ & $4.40$ & $3.83$ & $3.79$ & $6.04$ & $5.21$ &
$5.06$\\\hline
$%
\mathbb{Z}
_{2}^{(2)}$ & $4.40$ & $3.83$ & $3.79$ & $6.04$ & $5.21$ & $5.03$\\\hline
\end{tabular}
\ \ \ \ \ \ \ \ \ .
\end{equation}
Note that in both scenarios, there is a decrease in the values of the central
charges. The rather mild decrease provides further evidence that there is
only weak mixing between the probe and Standard Model.

Let us now compute the contribution to the SM beta functions.
Basically, this is dictated by the $T_{3}$ charge
assignments of those visible sector fields which have
developed a non-trivial scaling dimension in the IR theory. First consider the
$%
\mathbb{Z}
_{2}^{(1)}$ scenario. Here, we have:
\begin{align}
\delta b_{SU(3)}  &  =\frac{3k_{E_{8}}}{4}t-\frac{9}{2}\times\left(  \frac
{t}{2}+u_{10_{M}}\right) \\
\delta b_{SU(2)}  &  =\frac{3k_{E_{8}}}{4}t-\frac{9}{2}\times\left(  \frac
{t}{2}+u_{10_{M}}\right)  \,-\frac{3}{2}\times\left(  \frac{t}{2}+u_{H_{d}%
}\right) \\
\delta b_{U(1)}  &  =\frac{3k_{E_{8}}}{4}t-\frac{9}{2}\times\left(  \frac
{t}{2}+u_{10_{M}}\right)  -3\times\frac{18}{60}\left(  \frac{t}{2}+u_{H_{d}%
}\right)  .
\end{align}
Next consider the $%
\mathbb{Z}
_{2}^{(2)}$ scenario. Here, we have:%
\begin{align}
\delta b_{SU(3)}  &  =\frac{3k_{E_{8}}}{4}t  -\frac{9}{2}\times\left(  \frac{t}%
{2}+u_{10_{M}}\right) -\frac{3}{2} \times \left(  \frac{t}%
{2}+u_{\overline{5}_{M}}\right) \\
\delta b_{SU(2)}  &  =\frac{3k_{E_{8}}}{4}t -\frac{9}{2}\times\left(  \frac{t}%
{2}+u_{10_{M}}\right) -\frac{3}{2}\times\left(  \frac
{t}{2}+u_{\overline{5}_{M}}\right)  \,\\
\delta b_{U(1)}  &  =\frac{3k_{E_{8}}}{4}t  -\frac{9}{2}\times\left(  \frac{t}%
{2}+u_{10_{M}}\right) -\frac{3}{2}\times\left(  \frac
{t}{2}+u_{\overline{5}_{M}}\right)
\end{align}
which is a universal shift to the one-loop MSSM\ beta functions. Using the
expression $u=t-2/3$ and plugging in the value of $t_{\ast} \sim0.50$ yields:%
\begin{equation}%
\begin{tabular}
[c]{|c|c|c|c|}\hline
Beta Functions & $\delta b_{SU(3)}$ & $\delta b_{SU(2)}$ & $\delta b_{U(1)}%
$\\\hline
\multicolumn{1}{|c|}{$%
\mathbb{Z}
_{2}^{(1)}$} & $4.13$ & $4.$ & $4.05$\\\hline
\multicolumn{1}{|c|}{$%
\mathbb{Z}
_{2}^{(2)}$} & $4.$ & $4.$ & $4.$\\\hline
\end{tabular}
\ \ \ \ \ \ \ \ .
\end{equation}

\subsection{Summary of Fixed Points \label{sec:opbetafunc}}

Repeating a similar analysis as above for different monodromy choices, we can
find the anomalous dimensions for CFT and MSSM fields. From the perspective of
the CFT, the main thing we must specify is the $T_{3}$ charge of the various
operators. Below we summarize the operators and their corresponding $T_{3}$
charges, as well as the group-theoretic parameters $\mu_{1},\mu_{2}$ and $r$
entering into $R_{IR}$:%
\begin{equation}%
\begin{tabular}
[c]{|c|c|c|c|c|c|c|c|}\hline
$R_{IR}$\ Parameters & $\mu_{1}$ & $\mu_{2}$ & $r$ & $T_{3}(H_{u})$ &
$T_{3}\left(  H_{d}\right)  $ & $T_{3}\left(  \overline{5}_{M}\right)  $ &
$T_{3}\left(  10_{M}\right)  $\\\hline
$%
\mathbb{Z}
_{2}^{(1)}$ & $5/2$ & X & $1$ & $0$ & $-1/2$ & $0$ & $-1/2$\\\hline
$%
\mathbb{Z}
_{2}^{(2)}$ & $5/2$ & X & $1$ & $0$ & $0$ & $-1/2$ & $-1/2$\\\hline
$%
\mathbb{Z}
_{2}\times%
\mathbb{Z}
_{2}$ & $5/2$ & $5/2$ & $2$ & $0$ & $-1$ & $-1/2$ & $-1/2$\\\hline
$S_{3}$ & $7/2$ & $5/2$ & $4$ & $-1$ & $-1$ & $-1$ & $-1$\\\hline
$Dih_{4}^{(1)}$ & $9/2$ & $5/2$ & $10$ & $-2$ & $-3/2$ & $0$ & $-3/2$\\\hline
$Dih_{4}^{(2)}$ & $9/2$ & $5/2$ & $10$ & $0$ & $-3/2$ & $-2$ & $-3/2$\\\hline
\end{tabular}
\ \ \ \ \text{ \ \ \ \ }.
\end{equation}
An ``X'' indicates the entry of the table has no meaning.
See Appendix B for the definition of each monodromy scenario, and further
explanation of how the $T_{3}$ charge assignments are fixed for the modes
which dominantly couple to the probe D3-brane.

Given these values of the parameters, we can determine the form of the
infrared R-symmetry. For each scenario, those Standard Model fields which have
negative $T_{3}$ charge are those which are charged under $U(1)_{SM}$. Hence,
we can compute the cubic and linear anomalies, much as we did abstractly in
the previous sections, as well as in the explicit $%
\mathbb{Z}
_{2}$ monodromy scenarios described explicitly in the previous subsection.
Since the computation is rather similar in all cases, we shall omit the
details. Performing $a$-maximization, we can then extract various properties of
the IR theories. In all cases, the rather small shift in going from the
critical value of $t_{\Phi}$ of the intermediate $\Phi$-deformed theories,
to the eventual value of $t_{\ast}$ in the coupled probe/MSSM fixed point means that just as in
\cite{FCFT}, there are no obvious unitarity bound violations. We now turn to
the one and two D3-brane probe scenarios.

\subsubsection{Scenarios with One D3-Brane}

In this subsection we summarize the main computable properties of a single D3-brane probing a
particular configuration of seven-brane monodromy. To this end, we list the IR scaling dimensions of the
Higgs fields and third generation chiral matter fields, e.g. those which dominantly couple to the probe. We also
include the scaling dimensions of the corresponding $\mathcal{O}$ operators:
\begin{equation}
\begin{tabular}
[c]{|c|c|c|c|c|c|c|c|c|c|}\hline
IR Dimensions & $t_{\ast}$ & $H_{u}$ & $\mathcal{O}_{H_{u}}$ & $H_{d}$ &
$\mathcal{O}_{H_{d}}$ & $\overline{5}_{M}$ & $\mathcal{O}_{\overline{5}_{M}}$
& $10_{M}$ & $\mathcal{O}_{10_{M}}$\\\hline
$%
\mathbb{Z}
_{2}^{(1)}$ & $0.50$ & $1$ & $2.25$ & $1.13$ & $1.87$ & $1$ & $2.25$ & $1.13$
& $1.87$\\\hline
$%
\mathbb{Z}
_{2}^{(2)}$ & $0.50$ & $1$ & $2.25$ & $1$ & $2.25$ & $1.12$ & $1.88$ & $1.12$
& $1.88$\\\hline
$%
\mathbb{Z}
_{2}\times%
\mathbb{Z}
_{2}$ & $0.45$ & $1$ & $2.33$ & $1.35$ & $1.65$ & $1.01$ & $1.99$ & $1.01$ &
$1.99$\\\hline
$S_{3}$ & $0.36$ & $1.08$ & $1.92$ & $1.08$ & $1.92$ & $1.08$ & $1.92$ &
$1.08$ & $1.92$\\\hline
$Dih_{4}^{(1)}$ & $0.28$ & $1.25$ & $1.75$ & $1.04$ & $1.96$ & $1$ & $2.58$ &
$1.04$ & $1.96$\\\hline
$Dih_{4}^{(2)}$ & $0.27$ & $1$ & $2.59$ & $1.02$ & $1.98$ & $1.22$ & $1.78$ &
$1.02$ & $1.98$\\\hline
\end{tabular}
\end{equation}
In nearly all cases, we observe that there is only a
small shift in the scaling dimension of the Standard Model fields.

Additionally, we list the central charges $a$ and $c$ for the various
scenarios. To properly compare various theories, we include the contributions from $Z_1$, $Z_2$, $H_u$, $H_d$, $\overline{5}_{M}$ and $10_{M}$,
even if these modes are decoupled in the IR. Along these lines, we have included in $a_{UV}$ the contribution from the $\mathcal{N} = 2$ probe D3-brane,
as well as the decoupled chiral multiplets. Similar considerations apply for $c_{UV}$, $a_{\Phi}$ and $c_{\Phi}$:
\begin{equation}
\begin{tabular}
[c]{|c|c|c|c|c|c|c|}\hline
Central Charges & $a_{UV}$ & $a_{\Phi}$ & $a_{IR}$ & $c_{UV}$ & $c_{\Phi}$ &
$c_{IR}$\\\hline
$%
\mathbb{Z}
_{2}^{(1)}$ & $4.40$ & $3.83$ & $3.79$ & $6.04$ & $5.21$ & $5.06$\\\hline
$%
\mathbb{Z}
_{2}^{(2)}$ & $4.40$ & $3.83$ & $3.79$ & $6.04$ & $5.21$ & $5.03$\\\hline
$%
\mathbb{Z}
_{2}\times%
\mathbb{Z}
_{2}$ & $4.40$ & $3.50$ & $3.48$ & $6.04$ & $4.74$ & $4.66$\\\hline
$S_{3}$ & $4.40$ & $3.08$ & $3.05$ & $6.04$ & $4.19$ & $4.05$\\\hline
$Dih_{4}^{(1)}$ & $4.40$ & $2.49$ & $2.47$ & $6.04$ & $3.43$ & $3.35$\\\hline
$Dih_{4}^{(2)}$ & $4.40$ & $2.49$ & $2.45$ & $6.04$ & $3.43$ & $3.31$\\\hline
\end{tabular}
\end{equation}
In the above collection of numbers, the parameters $a_{\Phi}$ and $c_{\Phi}$
refer to \textquotedblleft intermediate\textquotedblright\ values of the
central charges given by the IR\ theory with all couplings to the Standard
Model switched off.

Finally, we also list the contributions to the beta functions for
the various scenarios. Here, the parameter $\delta b_{SU(5)}$ refers
to the contribution to the running of $SU(5)_{GUT}$ in the limit where the
F-term couplings to the Standard Model have been switched off, just as in equation (\ref{deltabsufiv}).
\begin{equation}
\begin{tabular}
[c]{|c|c|c|c|c|}\hline
Beta Functions & $\delta b_{SU(5)}$ & $\delta b_{SU(3)}$ & $\delta b_{SU(2)}$
& $\delta b_{U(1)}$\\\hline
$%
\mathbb{Z}
_{2}^{(1)}$ & $4.80$ & $4.13$ & $4.$ & $4.05$\\\hline
$%
\mathbb{Z}
_{2}^{(2)}$ & $4.80$ & $4.$ & $4.$ & $4.$\\\hline
$%
\mathbb{Z}
_{2}\times%
\mathbb{Z}
_{2}$ & $4.16$ & $4.00$ & $3.65$ & $3.79$\\\hline
$S_{3}$ & $3.50$ & $2.92$ & $2.75$ & $2.81$\\\hline
$Dih_{4}^{(1)}$ & $2.63$ & $2.38$ & $2.08$ & $2.20$\\\hline
$Dih_{4}^{(2)}$ & $2.63$ & $2.16$ & $2.14$ & $2.15$\\\hline
\end{tabular}
\ \ \ \ \ \text{ \ \ \ \ }.
\end{equation}
Note that the values of the beta functions are all consistent with the
requirement that $SU(5)_{GUT}$ remains asymptotically free. Further
note that in passing from the $\Phi$-deformed theory to the coupled probe/MSSM
system, there is only a small change in both the central charges, and the beta
functions. This is consistent with the expectation that the probe and Standard
Model only weakly mix.

\subsubsection{Scenarios with Two D3-Branes}

We can perform a similar analysis for the probe theories involving two
D3-branes. Our conventions are the same as in the case of a single D3-brane.
Basically, the only change is to the parameter $N$ entering into
the $\mathcal{N} = 2$ central charges $a_{E_{8}}$, $c_{E_{8}}$ and $k_{E_{8}}$.
The IR scaling dimensions are:
\begin{equation}
\begin{tabular}
[c]{|c|c|c|c|c|c|c|c|c|c|}\hline
IR Dimensions & $t_{\ast}$ & $H_{u}$ & $\mathcal{O}_{H_{u}}$ & $H_{d}$ &
$\mathcal{O}_{H_{d}}$ & $\overline{5}_{M}$ & $\mathcal{O}_{\overline{5}_{M}}$
& $10_{M}$ & $\mathcal{O}_{10_{M}}$\\\hline
$%
\mathbb{Z}
_{2}^{(1)}$ & $0.54$ & $1$ & $2.19$ & $1.22$ & $1.78$ & $1$ & $2.19$ & $1.22$
& $1.78$\\\hline
$%
\mathbb{Z}
_{2}^{(2)}$ & $0.54$ & $1$ & $2.19$ & $1$ & $2.19$ & $1.21$ & $1.79$ & $1.21$
& $1.79$\\\hline
$%
\mathbb{Z}
_{2}\times%
\mathbb{Z}
_{2}$ & $0.48$ & $1$ & $2.27$ & $1.45$ & $1.55$ & $1.09$ & $1.91$ & $1.09$ &
$1.91$\\\hline
$S_{3}$ & $0.40$ & $1.21$ & $1.79$ & $1.21$ & $1.79$ & $1.21$ & $1.79$ &
$1.21$ & $1.79$\\\hline
$Dih_{4}^{(1)}$ & $0.31$ & $1.41$ & $1.59$ & $1.17$ & $1.83$ & $1$ & $2.53$ &
$1.17$ & $1.83$\\\hline
$Dih_{4}^{(2)}$ & $0.31$ & $1$ & $2.54$ & $1.16$ & $1.84$ & $1.39$ & $1.61$ &
$1.16$ & $1.84$\\\hline
\end{tabular}
\end{equation}
We observe that these scaling dimensions are slightly larger than their single D3-brane counterparts. However,
the overall size is still on the small side.

Next consider the central charges of the UV, intermediate, and IR theories:
\begin{equation}
\begin{tabular}
[c]{|c|c|c|c|c|c|c|}\hline
Central Charges & $a_{UV}$ & $a_{\Phi}$ & $a_{IR}$ & $c_{UV}$ & $c_{\Phi}$ &
$c_{IR}$\\\hline
$%
\mathbb{Z}
_{2}^{(1)}$ & $11.4$ & $10.2$ & $10.1$ & $14.3$ & $12.7$ & $12.4$\\\hline
$%
\mathbb{Z}
_{2}^{(2)}$ & $11.4$ & $10.2$ & $10.1$ & $14.3$ & $12.7$ & $12.4$\\\hline
$%
\mathbb{Z}
_{2}\times%
\mathbb{Z}
_{2}$ & $11.4$ & $9.45$ & $9.39$ & $14.3$ & $11.7$ & $11.5$\\\hline
$S_{3}$ & $11.4$ & $8.43$ & $8.31$ & $14.3$ & $10.4$ & $10.0$\\\hline
$Dih_{4}^{(1)}$ & $11.4$ & $6.85$ & $6.77$ & $14.3$ & $8.46$ & $8.20$\\\hline
$Dih_{4}^{(2)}$ & $11.4$ & $6.85$ & $6.72$ & $14.3$ & $8.46$ & $8.11$\\\hline
\end{tabular}
\end{equation}
As expected, the central charge for the two D3-brane system is larger than that of the single D3-brane case. Note, however,
that the decrease in going from the $\Phi$-deformed theory to the coupled probe/MSSM system is still small.

Finally, we can also compute the values of the beta functions for these scenarios:
\begin{equation}
\begin{tabular}
[c]{|c|c|c|c|c|}\hline
Beta Functions & $\delta b_{SU(5)}$ & $\delta b_{SU(3)}$ & $\delta b_{SU(2)}$
& $\delta b_{U(1)}$\\\hline
$%
\mathbb{Z}
_{2}^{(1)}$ & $10.1$ & $9.10$ & $8.88$ & $8.96$\\\hline
$%
\mathbb{Z}
_{2}^{(2)}$ & $10.1$ & $8.84$ & $8.84$ & $8.84$\\\hline
$%
\mathbb{Z}
_{2}\times%
\mathbb{Z}
_{2}$ & $8.96$ & $8.36$ & $7.91$ & $8.09$\\\hline
$S_{3}$ & $7.70$ & $6.42$ & $6.$ & $6.17$\\\hline
$Dih_{4}^{(1)}$ & $5.90$ & $5.11$ & $4.53$ & $4.76$\\\hline
$Dih_{4}^{(2)}$ & $5.90$ & $4.69$ & $4.54$ & $4.60$\\\hline
\end{tabular}
\end{equation}
The values of the beta functions in most scenarios are significantly higher
than their single D3-brane counterparts. This leads to accelerated running of
the gauge couplings and typically a loss of asymptotic freedom for
$SU(5)_{GUT}$. Note, however, that in the \textquotedblleft large monodromy group
scenarios\textquotedblright\ such as the $S_{3}$ and the $Dih_{4}$ examples, the
effect is somewhat milder. A similar analysis can be performed for more than two
D3-branes. In all T-brane scenarios considered here, we lose asymptotic freedom for
$SU(5)_{GUT}$, so we do not entertain this possibility further.

\section{Threshold Corrections and Unification\label{sec:UNIFY}}

In the previous section we determined various properties of the fixed point
associated with coupling the MSSM to a strongly coupled CFT, and rather
surprisingly found that the effects on the Standard Model are mild. We have
also seen that coupling to the probe sector influences the running of the
gauge couplings, but in a way which is not $SU(5)_{GUT}$ universal. Indeed,
though the probe sector states fill out complete GUT multiplets, their
couplings to the Higgs sector explicitly break $SU(5)_{GUT}$. In this section
we discuss the consequences of such couplings for precision unification.

A compelling motivation for supersymmetric GUT theories is that at one loop order,
the gauge couplings of the MSSM appear to unify at a scale of order $\sim 10^{16}$ GeV
\cite{Dimopoulos:1981yj, Dimopoulos:1981zb, Ibanez:1981yh, Einhorn:1981sx, Marciano:1981un}. Beyond the
one-loop approximation, however, various effects can potentially distort unification. These distortions arise both
from low energy effects associated with two-loop contributions involving just the
MSSM\ degrees of freedom, as well as effects closer to the GUT scale.
Including only two-loop effects from the MSSM, it is
well-known that for typical superpartner masses,
if one runs the observed values of the gauge couplings to
higher energy scales without including any other threshold corrections, the
gauge coupling constants no longer unify. Defining $M_{GUT}$ as the energy
scale at which $\alpha_{1}(M_{GUT})=\alpha_{2}(M_{GUT})=\alpha_{GUT}$,
the value of $\alpha_{3}(M_{GUT})$ is lower than $\alpha_{GUT}$, and differs
from it by an order $4\%$ amount:%
\begin{equation}
\text{Two-Loop MSSM\ contribution: }\frac{\alpha_{3}^{-1}(M_{GUT}%
)-\alpha_{GUT}^{-1}}{\alpha_{GUT}^{-1}}\sim+4\%.
\end{equation}
See \cite{Raby:2006sk} for a recent review of such issues.

The precise amount of mismatch depends on the details of the superpartner mass
spectrum (see for example \cite{Langacker:1995fk, Raby:2009sf}). It is
equally well-known that various GUT scale threshold corrections from incomplete
GUT multiplets can induce an appropriate shift which can eliminate this
discrepancy (see for example
\cite{Ibanez:1991zv,Ibanez:1992hc,Nilles:1995kb,Hall:2001pg}). On the one
hand, this provides a way to retain gauge coupling unification. However, this
unification is achieved at the expense of including extra incomplete multiplets beyond the
Higgs fields. Let us note that such multiplets certainly exist in higher
dimensional theories such as F-theory GUTs, and are associated with the
Kaluza-Klein spectrum of excitations for the Higgs doublets and triplets.

Gauge coupling unification in F-theory GUTs has been studied in for example
\cite{DWII, Blumenhagen:2008aw, Conlon:2009qa, Marsano:2009wr, Leontaris:2009wi, Li:2010dp, Leontaris:2011pu}.
In the specific context of F-theory GUTs, a common way to break the GUT group
is through the introduction of hyperflux \cite{BHVII,DWII}. This introduces a
further distortion of gauge coupling unification, which is of the same order
and sign as the \textquotedblleft two-loop discrepancy\textquotedblright%
\ from purely MSSM\ effects \cite{DWII, Blumenhagen:2008aw}. For appropriate
values of the mass spectrum for the Higgs doublets and triplets, precision
unification can be retained \cite{DWII}. Though adding various thresholds from
incomplete GUT multiplets provides a potential way to rectify precision
unification, it is aesthetically displeasing that the least unified
parts of a Grand Unified Theory would somehow be destined to play the role of ensuring gauge
coupling unification.

In this section we show that the extra couplings between the probe sector
fields and the MSSM matter fields induce a correction to the running of the
gauge couplings which can counter these deleterious shifts to gauge
coupling unification. Moreover, this is achieved by adding a vector-like
sector with states which fill out complete GUT multiplets. The basic point,
however, is that because of the coupling to the Higgs sector, a shift is
induced in the running of the couplings.

At a qualitative level, the reason such couplings help with unification is due
to the ways that various distortions of gauge coupling unification enter. To
this end, it is helpful to recall that the numerator of the NSVZ beta function \cite{Novikov:1983uc} for a
gauge group $G$ is, in our sign conventions:
\begin{equation}
b_{G}^{NSVZ}=-3C_{2}(G)+\sum_{\Psi}C_{2}(R_{\Psi})(1-\gamma_{\Psi}).
\end{equation}
where $C_{2}(G)$ and $C_{2}(R_{\Psi})$ are the Dynkin indices
for the adjoint and representation $R_{\Psi}$, respectively. Further,
$\gamma_{\Psi}$ is the anomalous dimension associated with an elementary
field $\Psi$, and the sum is over the various elementary degrees of freedom of
the system.\footnote{In a strongly coupled non-Lagrangian theory there are
various subtleties associated with defining the elementary degrees of freedom.
In this paper we have bypassed this point by appealing to anomaly matching
considerations.} The main point is that the anomalous dimensions $\gamma$ of
the system also affect the running of the visible sector couplings. Indeed, the
two-loop distortion from MSSM\ effects comes from gauge
interactions. This contribution can be counteracted by
the effects of F-term couplings involving the Higgs fields. Such F-term couplings tend to
increase the anomalous dimensions of fields, which in turn modifies the form of the beta function.
As reviewed for example in \cite{Donkin:2010ta}, the top quark Yukawa is by itself not large enough to
correct the two-loop discrepancy generated by gauge interaction effects.
However, F-term couplings to additional sectors can lead to further shifts
which can counteract the two-loop MSSM\ discrepancy \cite{Donkin:2010ta}%
.\footnote{In \cite{Donkin:2010ta} it was assumed that the Higgs fields couple
to additional hidden sector fields through cubic terms involving two
additional matter fields charged under the Standard Model gauge group.
However, the qualitative effect is more general, and just requires the
existence of extra Yukawa couplings to the Higgs fields.} This is precisely
the situation we are in.

Here, we use our analysis of beta functions performed in the previous section
to estimate the size of threshold corrections to gauge coupling unification
induced by the probe D3-brane. Because there are various contributions to the
values of the GUT\ scale couplings, such as those induced by GUT\ scale
thresholds, e.g. hyperflux contributions and Kalua-Klein\ Higgs doublets and
triplets, as well as two-loop MSSM effects, we shall characterize all of these
effects in terms of a finite shift to the GUT\ scale values of the
inverse fine structure constants. From this perspective, precision unification
is achieved when the contribution from the one loop MSSM\ beta function and
the probe D3-brane threshold is:%
\begin{equation}
\text{Probe D3-brane\ contribution: }\frac{\alpha_{3}^{-1}(M_{GUT}%
)-\alpha_{GUT}^{-1}}{\alpha_{GUT}^{-1}}\sim- 4\%.
\end{equation}

Let us now turn to a more detailed discussion of threshold effects from the
probe sector. At some threshold scale $M_{thresh}\sim M_{\cancel{CFT}}$ we
assume that the additional CFT states enter. In a step function
approximation, the running of the various gauge couplings are:%
\begin{align}
\alpha_{3}^{-1}(l)  &  =\left(-  \frac{b_{SU(3)}^{(0)}}{2\pi}\cdot
(l-l_{0})+\alpha_{3}^{-1}(l_{0})\right)  \times\theta\left(  l_{t}-l\right)
\\
&  +\left(-  \frac{b_{SU(3)}^{(0)} + \delta b_{SU(3)}}{2\pi}\cdot(l-l_{t})-\frac
{b_{SU(3)}^{(0)}}{2\pi}\cdot(l_{t} - l_{0})+\alpha_{3}^{-1}(l_{0})\right)
\times\theta\left(  l-l_{t}\right) \\
\alpha_{2}^{-1}(l)  &  =\left(-  \frac{b_{SU(2)}^{(0)}}{2\pi}\cdot
(l-l_{0})+\alpha_{2}^{-1}(l_{0})\right)  \times\theta\left(  l_{t}-l\right)
\\
&  +\left(-  \frac{b_{SU(2)}^{(0)} + \delta b_{SU(2)}}{2\pi}\cdot(l-l_{t})-\frac
{b_{SU(2)}^{(0)}}{2\pi}\cdot(l_{t} - l_{0})+\alpha_{2}^{-1}(l_{0})\right)
\times\theta\left(  l-l_{t}\right) \\
\alpha_{1}^{-1}(l)  &  =\left(-  \frac{b_{U(1)}^{(0)}}{2\pi}\cdot
(l-l_{0})+\alpha_{1}^{-1}(l_{0})\right)  \times\theta\left(  l_{t}-l\right) \\
&  +\left(-  \frac{b_{U(1)}^{(0)} + \delta b_{U(1)}}{2\pi}\cdot(l-l_{t})-\frac{b_{U(1)}%
^{(0)}}{2\pi}\cdot(l_{t} - l_{0})+\alpha_{1}^{-1}(l_{0})\right)  \times
\theta\left(  l-l_{t}\right)
\end{align}
where $l>l_{0} > 0$ denotes the RG time of the evolution, $l_{0}$ is a reference
value for the entry of the superparticles, and $l_{t}$ is the RG time of
the threshold. Here, $b^{(0)}_{G}$ refers to the usual one-loop MSSM beta functions,
and $\delta b_{G}$ refers to the threshold correction induced from coupling to the probe
D3-brane.

In the previous section we have seen that the beta function contributions $\delta b_{G}$
are not $SU(5)_{GUT}$ universal. We find that this leads to a distortion of unification
of the same size as two loop MSSM effects, but in the \textit{opposite} direction.
The contribution from such probe sectors therefore can actually help with precision unification.

To study the overall size of these effects, we now consider some representative
values for the threshold scale $M_{thresh} \sim M_{\cancel{CFT}}$. In principle,
we should also include the effects of the precise mass scales for all
superpartners. Rather than entangling this effect with the contribution from
just the probe sector, we simply take $l_{0} \sim 500$ GeV, with the
gauge couplings at the first threshold scale to be:%
\begin{equation}
(\alpha_{3}^{-1}(l_{0}),\alpha_{2}^{-1}(l_{0}),\alpha_{1}%
^{-1}(l_{0}))\sim(10,30,58).
\end{equation}
We note that with these values, the one-loop running
leads to a unified value of $\alpha_{GUT} \sim 0.05$ at
a scale $M_{GUT} \sim 2 \times 10^{16}$ GeV.

We define the GUT scale to be the scale at which
$\alpha_{2}^{-1}$ and $\alpha_{1}^{-1}$ unify, and we refer to this unified value
as $\alpha_{GUT}^{-1}$. We denote the percent mismatch between $\alpha_{3}$ and $\alpha_{GUT}$ by:%
\begin{equation}
\delta\equiv\frac{\alpha_{3}^{-1}(M_{GUT})-\alpha_{GUT}^{-1}}{\alpha
_{GUT}^{-1}}.
\end{equation}
For a threshold scale $M_{thresh} \sim 10^{13}$ GeV, this yields the
following numerical values for the various monodromy scenarios:
\begin{equation}%
\begin{tabular}
[c]{|c|c|c|c|c|c|c|c|}\hline
$M_{thresh}\sim10^{13}$ GeV & $\delta b_{SU(3)}$ & $\delta b_{SU(2)}$ &
$\delta b_{U(1)}$ & $M_{GUT}$ (GeV) & $\alpha_{GUT}^{-1}$ & $\alpha_{3}^{-1}$
& $\delta$\\\hline
$%
\mathbb{Z}
_{2}^{(1)}$ & \multicolumn{1}{|c|}{$4.13$} & $4.$ & $4.05$ & $2\times10^{16}$
& $20.2$ & $20.0$ & $-1\%$\\\hline
$%
\mathbb{Z}
_{2}^{(2)}$ & \multicolumn{1}{|c|}{$4.$} & $4.$ & $4.$ & $2\times10^{16}$ &
$20.1$ & $20.1$ & $0\%$\\\hline
$%
\mathbb{Z}
_{2}\times%
\mathbb{Z}
_{2}$ & $4.00$ & $3.65$ & $3.79$ & $2\times10^{16}$ & $20.7$ & $20.1$ &
$-3\%$\\\hline
$S_{3}$ & \multicolumn{1}{|c|}{$2.92$} & $2.75$ & $2.81$ & $2\times10^{16}$ &
$21.7$ & $21.4$ & $-1\%$\\\hline
$Dih_{4}^{(1)}$ & \multicolumn{1}{|c|}{$2.38$} & $2.08$ & $2.20$ &
$2\times10^{16}$ & $22.5$ & $22.1$ & $-2\%$\\\hline
$Dih_{4}^{(2)}$ & $2.16$ & $2.14$ & $2.15$ & $2\times10^{16}$ & $22.4$ &
$22.3$ & $-0.1\%$\\\hline
\end{tabular}
\ \ \ \ \ \ \ \ \ \ \ \ \ .
\end{equation}
From this, we see that in the one case where the beta functions are the same,
we retain one-loop unification, while in those cases with a small shift, there
is an improved agreement with precision unification. Of course, the value of $M_{GUT}$ quoted here does not include
other TeV and GUT scale thresholds, so this should really be taken as an order of magnitude estimate.

It is also of interest to compute the size of GUT distorting effects when the
threshold scale is pushed down to $\sim500$ GeV, which is close to the
maximal value allowed by present bounds. This is an interesting possibility to
see the largest possible distortion of gauge coupling unification in the
presence of such a probe sector. In this case, we obtain:%
\begin{equation}%
\begin{tabular}
[c]{|c|c|c|c|c|c|c|c|}\hline
$M_{thresh}\sim500$ GeV & $\delta b_{SU(3)}$ & $\delta b_{SU(2)}$ & $\delta
b_{U(1)}$ & $M_{GUT}$ (GeV) & $\alpha_{GUT}^{-1}$ & $\alpha_{3}^{-1}$ &
$\delta$\\\hline
$%
\mathbb{Z}
_{2}^{(1)}$ & $4.13$ & $4.$ & $4.05$ & $2\times10^{16}$ & $5.2$ & $4.4$ &
$-16\%$\\\hline
$%
\mathbb{Z}
_{2}^{(2)}$ & $4.$ & $4.$ & $4.$ & $2\times10^{16}$ & $5.0$ & $5.0$ &
$0\%$\\\hline
$%
\mathbb{Z}
_{2}\times%
\mathbb{Z}
_{2}$ & $4.00$ & $3.65$ & $3.79$ & $1\times10^{16}$ & $7.3$ & $5.1$ &
$-30\%$\\\hline
$S_{3}$ & $2.92$ & $2.75$ & $2.81$ & $2\times10^{16}$ & $11.5$ & $10.4$ &
$-9\%$\\\hline
$Dih_{4}^{(1)}$ & $2.38$ & $2.08$ & $2.20$ & $1\times10^{16}$ & $14.9$ &
$13.1$ & $-12\%$\\\hline
$Dih_{4}^{(2)}$ & $2.16$ & $2.14$ & $2.15$ & $2\times10^{16}$ & $14.3$ &
$14.2$ & $-1\%$\\\hline
\end{tabular}
\ \ \ \ \ \ \ \ \ \ \ \ \ \ \ \ \ \ \ \ \ .
\end{equation}
As in the higher threshold case, the value of $M_{GUT}$ should really only be viewed as an order of magnitude estimate
since here we are neglecting various TeV and GUT scale thresholds. In this case,
the effects of the probe D3-brane sometimes induce a significant
\textquotedblleft overshoot\textquotedblright\ in the value of $\alpha_{3}$
versus $\alpha_{GUT}$ in the direction opposite to that expected from the
two-loop MSSM\ effects. It is interesting to note that even when we push the
scale of CFT\ breaking as low as $500$ GeV, the $S_{3}$ and $Dih^{(1)}_{4}$ monodromy scenarios
induce only an order $10 \%$ distortion in gauge coupling unification, while for the $Dih^{(2)}_{4}$ scenario,
the amount of distortion is on the order of $1 \%$.

Of course, to perform a more complete analysis of precision unification, we
should work to two loop order in the MSSM\ gauge couplings, include the
corresponding threshold corrections, and include all GUT scale
threshold corrections as well. This would involve a more detailed analysis
beyond the scope of the present paper. Nevertheless, the qualitative effect,
as well as its overall size, is clear: The presence of the probe D3-brane
provides all the necessary ingredients to improve precision unification.

\section{Supersymmetry Breaking\label{sec:SUSY}}

In this section we briefly consider the possibility that the hidden sector can
serve as the origin of supersymmetry breaking. In particular we argue that it
is natural in this setup to obtain a deformation of gauge-mediated
supersymmetry breaking. As discussed earlier, the breaking of the CFT is
naturally implemented by moving the D3-brane off the $7_{SM}$-brane. This displacement corresponds
to giving a vev to the operator $Z$ of the probe theory. We will also argue
that this procedure provides a natural mechanism for breaking supersymmetry. Once
the D3-brane moves off the $7_{SM}$-brane, the $3-7_{SM}$ strings which are
charged under the SM gauge group will serve as messengers in a gauge
mediated scenario, with the additional novelty that
the effective ``number'' of messenger fields is generically irrational but
computable. In addition, as already noted, the geometric profile of matter
fields in the internal directions of the compactification singles out the
Higgs fields and third generation as the dominant sources of coupling between
the probe and Standard Model. This leads to additional supersymmetry breaking
effects for the third generation and Higgs sectors, which can induce
potentially significant contributions to the $\mu,B\mu$, $A$-terms, and soft masses.

The organization of the rest of this section is as follows. First, we explain
why it is natural to expect supersymmetry breaking to occur in the probe brane
setup. Next, we discuss the resulting contributions from gauge mediated
supersymmetry breaking, and after this, the expected deformations to the Higgs
fields and third generation fields. Finally, we discuss how various
considerations from combining the probe and Standard Model sectors suggest
natural ranges of energy scales for the messenger scale of the model.

\subsection{CFT and SUSY Breaking on the D3-Brane}

In this subsection we discuss a geometrically natural way to implement
both conformal symmetry breaking and supersymmetry breaking on a probe D3-brane. The
basic scenario is as follows. We imagine that in the vicinity of the
configuration of intersecting seven-branes which realizes the Standard Model,
there is nearby a D3-brane. The position of this D3-brane in the geometry is
stabilized by various fluxes \cite{Martucci, FGUTSNC, Baumann:2010sx}. The general form of the
flux-induced superpotential will be a power series in the modes $Z_{i}$:%
\begin{equation}
W_{flux}=F_{i}Z_{i}+m_{ab}Z_{a}Z_{b}+\lambda_{ijk}Z_{i}Z_{j}Z_{k}+O(Z^{4}).
\end{equation}
The precise form of $W_{flux}$ will not concern us here. This deformation
persists even in the limit of non-compact seven-branes, because it can be
stated as a deformation of the topological B-model, independent of K\"{a}hler
data \cite{FGUTSNC}. Even so, the actual size of the deformation will depend on such data
since $W_{flux}$ is more accurately thought of as a section of a bundle. Once this
extra data is fixed, we see that $W_{flux}$ becomes quite dilute as we
decompactify the seven-brane (see also \cite{Marchesano:2009rz}). It is also
technically natural to treat $W_{flux}$ as a small perturbation to the
system. This is because $W_{flux}$ breaks additional flavor symmetries
of the probe sector. In what follows, we shall therefore ignore the
effects of $W_{flux}$ in determining the IR behavior of our probe theory.
Let us note that the presence of $W_{flux}$ will induce a further deformation of the
conformal sector. In \cite{FCFT} it was found that terms linear in $Z_{1}$ and
$Z_{2}$ do not lead to an interacting CFT in the infrared. See Appendix B of
\cite{D3gen} for a discussion of the fixed points obtained from terms
quadratic in the $Z_{i}$.

Thinking of $W_{flux}$ as a small perturbation, the presence of terms in $Z$ will
tend to attract the D3-brane close to the Yukawa point. Indeed, the fact that
such terms are irrelevant deformations of the CFT means that there will be a
supersymmetric vacuum located at $Z=0$, the origin of the Coulomb branch. For
generic enough $W_{flux}$, however, we can expect additional vacua close to the
Yukawa point. Indeed, when generic mass deformations for the $\mathcal{N}=2$
theory are switched on, for any point along the Coulomb branch of moduli
space there exists a suitable choice of $W_{flux}$ such that a metastable
vacuum at that prescribed value can be found \cite{Ooguri:2007iu, Auzzi:2010kv}. In our context with
$\mathcal{N}=1$ vacua, it is natural to expect that a similar situation occurs.

When $Z$ develops a non-zero vev, the D3-brane moves out on the Coulomb
branch, and conformal symmetry is also broken. Let us denote by
$M_{\cancel{CFT}}$ the energy scale at which conformal symmetry is broken.
Heuristically, we identify $\left\langle Z\right\rangle ^{1/\Delta}\sim
M_{\cancel{CFT}}$. We can therefore promote the effects of the vev to a
spurion chiral superfield $X$, which has scalar vev $M_{\cancel{CFT}}$.
Phrased in this way, we can also include the effects of supersymmetry breaking
in terms of the vev:%
\begin{equation}
\left\langle X\right\rangle =M_{\cancel{CFT}}+\theta^{2}F.
\end{equation}

Because $M_{\cancel{CFT}}$ specifies the scale of conformal symmetry breaking,
it follows that below this energy scale, the dynamics of the probe theory is
described by a theory with particle-like excitations. Note, however, that
since $\tau\sim O(1)$, this will be a strongly coupled theory. This
characteristic energy scale specifies the masses of some of the particle-like
excitations. For example, the mediator $3-7_{SM}$ strings, namely states
charged under the Standard Model gauge group will have mass $M_{\cancel{CFT}}%
$:%
\begin{equation}
M_{mess}\sim M_{\cancel{CFT}}.
\end{equation}

An interesting feature of probes with seven-brane monodromy is that some of
the GUT singlets of the probe sector will have masses far below
$M_{\cancel{CFT}}$, given instead by \cite{D3gen}:
\begin{equation}
M_{hid}\sim M_{\cancel{CFT}}\left(  \frac{M_{\cancel{CFT}}}{M_{GUT}}\right)
^{\alpha}%
\end{equation}
for $\alpha\sim O(1)$ set by the details of seven-brane
monodromy.\footnote{For example, in weakly coupled models such as the $D_{4}$
probe theory $\alpha=1,2,3$, with the value set by the details of seven-brane
monodromy. Similar considerations are expected to hold in general
\cite{D3gen}.} Of course, the exact spectrum will also depend on various
D-terms and possible supersymmetry breaking effects.

\subsection{Gauge Mediated Contributions}

Introducing a source of supersymmetry breaking through the $Z$ field produces a
non-supersymmetric mass spectrum for the $3-7_{SM}$ strings. The fact that these
states are charged under the Standard Model gauge group means that the effects
of supersymmetry breaking in the probe sector will be communicated via gauge
interactions to the visible sector. In other words, there is a natural mode of SUSY breaking transmission
via gauge mediation.

To illustrate the basic ideas, we imagine that $Z$ develops a supersymmetry
breaking vev:%
\begin{equation}
\left\langle Z\right\rangle =\left(  M_{mess}+\theta^{2}F\right)
^{\Delta_{IR}(Z)}.
\end{equation}
We now show that in the limit $F/M_{mess}^{2}\ll1$, it is possible to extract
the soft supersymmetry breaking contributions from gauge mediated
supersymmetry breaking. Our discussion closely follows \cite{Giudice:1997ni}.

To this end, recall that we have computed the shift in the SM gauge coupling
beta functions expected from coupling to the probe sector. Indeed, for%
\begin{equation}
b_{G}=-3\,\mathrm{Tr}(R_{IR}J_{G}J_{G}),
\end{equation}
with $G=SU(3),SU(2),U(1)$, we have defined the contribution from the messenger
sector to be given by a threshold contribution:%
\begin{equation}
\delta b_{G}=b_{G}-b_{G}^{(0)}.
\end{equation}
Let us now suppose that the $3-7_{SM}$ strings pick up a supersymmetric mass
$M_{mess}$. This is the messenger scale for our model. There is a Lagrangian
F-term contribution of the form
\begin{equation}
\delta\mathcal{L}=\frac{1}{8\pi}\operatorname{Im}\int d^{2}\theta\,\tau
_{G}\,\mathrm{Tr}_{G}\mathcal{W}^{\alpha}\mathcal{W}_{\alpha}-\frac{\delta
b_{G}}{16\pi^{2}}\operatorname{Re}\int d^{2}\theta\log M_{mess}\,\mathrm{Tr}%
_{G}\mathcal{W}^{\alpha}\mathcal{W}_{\alpha}. \label{running}%
\end{equation}
If we now imagine that the $3-7_{SM}$ mass spectrum is non-supersymmetric
$M_{mess}\rightarrow M_{mess}+\theta^{2}F$, we obtain a gaugino mass at the messenger scale:
\begin{equation}
m_{\lambda}=\delta b_{G}\left(  \frac{\alpha_{G}}{4\pi}\right)  \left(
\frac{F}{M_{mess}}\right)  .
\end{equation}
Thus the effective number of messengers $N_{G}$ is, as expected, simply
\begin{equation}
N_{G}=\delta b_{G}.
\end{equation}
The interesting point in our case is that this number will generically be
irrational because of the CFT dynamics.

For the scalars there is the usual
\textquotedblleft two-loop" contribution from gauge mediation, where one of
the loops stands in for the CFT dynamics. As with the gauginos, it
is simple to read off the gauge mediated
contribution from the running of the gauge coupling. At the messenger scale,
this soft mass term is given by:
\begin{equation}
m_{\tilde{f}}^{2}=2\sum_{G}C^{G}_{2}(R_{\tilde{f}})N_{G}\left(  \frac{\alpha_{G}%
}{4\pi}\right)  ^{2}\left(  \frac{F}{M_{mess}}\right)  ^{2}%
\end{equation}
As before, this is what we expect as the contribution from $N_{G}$ messengers
for each gauge group $G$.

\subsection{Deformation of Gauge Mediation}

As noted before, there are additional couplings between the probe and the
visible sector. Geometrically, the main effects are from those Standard Model
fields which have maximal wavefunction profile at the point where the D3-brane
sits. In the flavor physics scenario of \cite{HVCKM, BHSV, FGUTSNC}, this
means there is a dominant coupling to the third generation and the Higgs fields.

Indeed, at the GUT scale, the Standard Model and probe become coupled via the
F-term deformation:
\begin{equation}
\delta W_{SM\oplus D3}=H_{u}\mathcal{O}_{H_{u}}+H_{d}\mathcal{O}_{H_{d}%
}+\overline{5}_{M}\mathcal{O}_{\overline{5}_{M}}+10_{M}\mathcal{O}_{10_{M}.}
\label{mixmix}%
\end{equation}
We can view the $\mathcal{O}$'s, which are operators in the CFT, as roughly
being made of composites of messenger fields charged under the MSSM gauge
group, whose beta function contributions we have already discussed. Integrating
these modes out below the $M_{\cancel{CFT}}$ scale will induce higher
dimension operators involving MSSM singlet operators from the D3-brane
sector coupled to the MSSM fields. In particular we generate an F-term
\begin{equation}
W_{eff}\supset\int d^{2}\theta\text{ }\mathcal{O}_{PQ}H_{u}H_{d}%
\end{equation}
where $\mathcal{O}_{PQ}$ is an MSSM neutral operator of the D3 brane theory
charged under the Peccei-Quinn symmetry. Here, we implicitly view this as a
higher dimension operator with suppression scale set by the mass scale of the
heavy messenger states, which is $M_{\cancel{CFT}}$. This term can provide a
mechanism for generating $\mu$ and $B\mu$ terms if $\mathcal{O}_{PQ}$ picks up
a vev. One can imagine that the $3-7_{SM}$ strings also communicate supersymmetry breaking
to the hidden sector via gauge mediated effects involving the $U(1)_{D3}$ of the probe theory.
In particular, various radiative corrections could naturally induce the analogue of
\textquotedblleft electroweak symmetry breaking\textquotedblright\ in the
hidden sector. Viewing $\mathcal{O}_{PQ}$ as a composite of $7_{flav}-3$ and
$3-7_{flav}$ modes, the basic idea is that these $3-7_{flav}$ strings can then
condense with supersymmetry breaking vevs which also break the PQ symmetry,
thus inducing $\mu$ and $B\mu$ terms. We are currently studying the dynamics
associated with such a scenario \cite{HVW}.

D-term corrections will also be generated. Some such corrections are already
present from gauge mediated contributions. Indeed, in minimal gauge mediation,
the soft scalar masses can be computed by tracking the dependence of
wave-function renormalization factors on a supersymmetry breaking spurion
\cite{Giudice:1997ni, ArkaniHamed:1998kj}. The presence of the F-term couplings
$\Psi_{R} \cdot \mathcal{O}_{\overline{R}}$ introduce additional messenger/matter couplings,
involving mainly the third generation and Higgs fields. These introduce
additional contributions to the soft masses and $A$-terms. Let us note that
because the dominant coupling is to the third generation, we do not expect
such contributions to induce large flavor changing neutral currents in the
first two generations.
%
%
As one can see from this discussion, there is a rich set of possibilities
for potential soft supersymmetry breaking patterns motivated by D3-brane considerations.

\subsection{Comments on the CFT Breaking/Messenger Scale}

Up to this point, we have treated the CFT breaking scale (which is the
messenger scale) as an undetermined parameter. There are two considerations
which allow us to narrow down the values of this parameter. First, we observe
that if we go all the way to the infrared fixed point, then the Standard Model
Yukawa couplings involving the ``third generation'' are all irrelevant. In this case, it would be
more appropriate to view the mode identified at the GUT scale as a ``second generation'' up type quark as
an effectively ``third generation'' field with a tuned top quark Yukawa. We do not entertain this possibility here.

Keeping the CFT breaking scale as low as possible is appealing in the gauge mediated framework, because the CFT
breaking and supersymmetry breaking scales can then be roughly the same.
To see how low we can push the CFT breaking scale, let us consider the amount of suppression expected for the top quark Yukawa coupling
in running from the GUT scale down to the CFT breaking scale. To illustrate the main idea, suppose
that the dimension of the $5_{H} \times 10_{M} \times 10_{M}$ term is
of order $3 + \epsilon \sim 3.1$, which is numerically similar to the values obtained in many of the scenarios.
In this case, demanding that the Yukawa coupling is sufficiently large imposes
the constraint:
\begin{equation}
\lambda_{5 \times 10 \times 10}  \sim\left(  \frac{M_{\cancel{CFT}}}{M_{GUT}}\right)^{0.1} \gtrsim 10^{-1}
\end{equation}
where here, we allow an order $10$ fine-tuning in the GUT scale value of the Yukawa.
This leads to a lower bound on the CFT breaking scale, of order:
\begin{equation}
M_{\cancel{CFT}}\gtrsim10^{3}\,\,\text{TeV.}%
\end{equation}
While this is close to the TeV scale, it is clearly above it. The precise lower bound depends on
the amount of tuning of the Yukawas at the GUT scale, and the particular probe scenario in question.
In other words, considerations from the visible sector, in particular that flavor physics
be generated correctly at the GUT scale, disfavors TeV scale values for the
hidden sector.

Confining our attention to the most conservative possibility where flavor is generated
at the GUT scale, retaining sufficiently large Yukawas then suggests increasing
the CFT breaking scale to an intermediate value, of order $10^{9} - 10^{13}$ GeV. This
is also in the range which is most helpful for precision unification in most
of the scenarios we have considered. Additionally, this is the range of values
which are most natural for D3-brane induced baryon and dark matter genesis
scenarios \cite{D3gen}. In this regime, there is also a
natural hierarchy between the mass scale of the messenger particles, and the
mass scale of the GUT singlets \cite{D3gen}. This fits nicely with the fact that
if the $\mu$-term is generated at lower energy scales, it should involve the
dynamics of light degrees of freedom below the CFT breaking scale.

Let us note that especially in the single D3-brane $Dih^{(2)}_{4}$ probe
theory, the amount of suppression to the Yukawas is quite mild. In this case, the dimension of the $5_{H} \times 10_{M} \times 10_{M}$
operator is $\sim 3 + 0.04$, while that of the $\overline{5}_{H} \times \overline{5}_{M} \times 10_{M}$ has dimension of order
$\sim 3 + 0.26$. Since the bottom quark is significantly lighter than the top quark,
one could also contemplate a moderate to low $\tan \beta$ scenario in which the bottom quark
is instead identified with the GUT scale ``second generation'' mode. Indeed, in the
flavor scenario of \cite{HVCKM}, the ratio of the second and third generation masses at the GUT scale is only
of order $\alpha_{GUT} \sim 0.05$. This might allow significantly lower CFT breaking scales
for the $Dih^{(2)}_{4}$ scenario, which in principle could be accessible at the LHC.
Note that the amount of distortion to unification in the $Dih^{(2)}_{4}$ scenario is also smaller than
in the other cases, and begins to counteract two-loop MSSM effects when the CFT breaking scale is quite low.
Of course, this would also require us to revisit various assumptions about flavor physics. Though a full study
of such a scenario is beyond the scope of the present work, this is perhaps the most phenomenologically exciting possibility.

\section{Conclusions \label{sec:CONC}}

E-type Yukawa points are required in order to generate a top quark mass in
F-theory GUTs. D3-brane probes of such E-points are then a very well-motivated
extension of the Standard Model. In this paper we
have studied the effects of the Standard Model on such a probe sector, and
conversely, the effects of the probe on the Standard Model. We have presented
evidence for the existence of a strongly interacting conformal fixed point for
this system, and moreover, have shown that various properties of this system,
such as the infrared R-symmetry, the scaling dimensions of operators, and the
effects of the probe on the running of the gauge coupling constants are all
computable. This analysis reveals some remarkable features, notably:
\begin{itemize}
\item Such fixed points exist.
\item The MSSM fields develop
small anomalous dimensions, and thus maintain their identity in the IR.
\item The presence of the probe sector states coming in complete GUT
multiplets can help with precision unification.
\end{itemize}
We have also seen that the
D3-brane can serve as a natural source of supersymmetry breaking in
which the gauginos and first two generations have a spectrum similar to gauge
mediated supersymmetry breaking, with additional deformations of the
third generation and Higgs sectors. In the remainder of this section we discuss
various future directions.

In this paper we have mainly focused on the regime of energy scales
between the CFT breaking scale $M_{\cancel{CFT}}$ and the GUT scale. Below
this scale, various fields will develop vevs, and they will affect the low
energy phenomenology. Some studies of metastable vacua for theories with
underlying $\mathcal{N} = 2$ supersymmetry have been considered in the
literature. It would clearly be of interest to extend this analysis to the
types of systems studied here. Along these lines, it is also natural to
consider how supersymmetry breaking on the D3-brane would influence the
visible sector. We have seen that the probe D3-brane system leads to a pattern
of supersymmetry breaking terms which mainly deforms the third generation and
Higgs fields away from what is expected in minimal gauge mediation.
Determining the full range of possibilities, and the associated phenomenology
would clearly be of interest.

We have also seen that in spite of appearances, qualitative as well
as quantitative features of the probe/MSSM couplings can be extracted from this
scenario. In particular, we have studied the running of the gauge couplings,
and the consequences for gauge coupling unification. After moving onto the
Coulomb branch, the probe D3-brane contributes another $U(1)_{D3}$ gauge group
factor, which is strongly coupled at the CFT breaking scale. In particular,
the value of the holomorphic gauge coupling $\tau_{D3}(M_{\cancel{CFT}})
\sim\exp(2 \pi i /3)$ near an $E_{6}$ or $E_{8}$ point. At lower scales, this
$U(1)_{D3}$ is expected to be broken. It would be interesting to see whether an
exact computation of the running down to the $U(1)_{D3}$ breaking scale could also
be performed. It would also be of interest to determine the exact type of
kinetic mixing expected between the abelian factor of the Standard Model and
this $U(1)_{D3}$ gauge group factor.

Another feature of the probe sector is that it contains operators with the
same gauge quantum numbers as the usual MSSM Higgs fields. Most of the usual
headaches with electroweak symmetry breaking and its naturalness stem from
the vector-like nature of the Higgs fields. Though somewhat removed from the
considerations of the present paper, an intriguing possibility is that the
operators of the probe sector could effectively function as Higgs fields. This
can be interpreted as dissolving the D3-brane into a finite size instanton in
the visible sector seven-brane.

\newpage

\section*{Acknowledgements}

We thank N. Arkani-Hamed, C. C\'{o}rdova, D. Green, Z. Komargodski, P. Langacker, D. Poland, and S.-J. Rey for helpful
discussions. JJH and BW thank the Harvard high energy theory group for
generous hospitality during part of this work. CV thanks the CTP at MIT for
hospitality during his sabbatical leave. The work of JJH is supported by NSF
grant PHY-0969448. The work of CV is supported by NSF grant PHY-0244821. The
work of BW is supported by the US Department of Energy under grant DE-FG02-95ER40899.


\appendix

\section{Probing T-Branes}

In this Appendix we briefly review how to analyze the matter content of
T-brane configurations, and how this matter couples to probe D3-branes. Our
discussion follows that given in \cite{TBRANES}. For further discussion on the computation of localized zero modes
in T-brane configurations and the associated Yukawa couplings, see \cite{TBRANES, Chiou:2011js, CordovaToAppear}.

We begin by discussing the visible sector matter fields in holomorphic gauge.
The basic idea is that the matter fields of the visible sector are organized
according to the decomposition of the adjoint of $E_{8}$ into irreducible
representations of $SU(5)_{GUT}\times SU(5)_{\bot}$:%
\begin{align}
E_{8}  &  \supset SU(5)_{GUT}\times SU(5)_{\bot}\\
248  &  \rightarrow(1,24)+(24,1)+(10,5)+(\overline{10},\overline
{5})+(5,\overline{10})+(\overline{5},10).
\end{align}
The zero mode content is specified by a holomorphic cohomology theory which
involves acting by $\Phi$ on appropriate representations of $SU(5)_{\bot}$.
In holomorphic gauge, a matter field $\Psi$ which descends from the
representation $(R_{GUT},R_{\bot})$ is given by an expression of the form:%
\begin{equation}
\Psi=\Phi\cdot\xi+h
\end{equation}
for holomorphic $\xi$ and $h$. The space of $h$'s modulo holomorphic gauge
transformations, e.g. contributions of the form $\Phi\cdot\xi$, define the zero
mode content of the theory.

To illustrate the action of $\Phi$, we introduce a basis of vectors
$e_{1},...,e_{5}$ which span the $5$ of $SU(5)_{\bot}$. We can also introduce
a basis for the $10$ of $SU(5)_{\bot}$ given by $e_{i}\wedge e_{j}$. The
action of $\Phi$ on $e_{i}\wedge e_{j}$ is specified in terms of its action on
the fundamental representation:%
\begin{equation}
\Phi\cdot\left(  e_{i}\wedge e_{j}\right)  =\left(  \Phi\cdot e_{i}\right)
\wedge e_{j}+e_{i}\wedge\left(  \Phi\cdot e_{j}\right)  .
\end{equation}
For example, consider a $%
\mathbb{Z}
_{2}$ monodromy scenario with:%
\begin{equation}
\Phi=\left[
\begin{array}
[c]{ccccc}%
a & 1 &  &  & \\
Z_{1} & a &  &  & \\
&  & b &  & \\
&  &  & c & \\
&  &  &  & d
\end{array}
\right]
\end{equation}
and $a,b,c,d$ linear expressions in the coordinates $Z_{1}$ and $Z_{2}$
subject to the constraint $2a+b+c+d=0$. To work out the zero mode content, we
let $\Phi$ act on a five-component vector $v=v_{i}e_{i}$:
\begin{equation}
\Phi\cdot v=\left[
\begin{array}
[c]{c}%
av_{1}+v_{2}\\
Z_{1}v_{1}+av_{2}\\
bv_{3}\\
cv_{4}\\
dv_{5}%
\end{array}
\right]  .
\end{equation}
One can now specify a gauge for which the top
component of the zero mode vector $\Psi$ vanishes. This leaves us with four
localized zero modes at $Z_{1}=a^{2}$, $b=0$, $c=0$ and $d=0$.

One way to crudely characterize such localized modes is in terms of the action
of a \textquotedblleft monodromy group\textquotedblright\ \cite{Hayashi:2009ge, BHSV, Marsano:2009gv}.
In mathematical terms, this is the Galois group associated with the characteristic polynomial for $\Phi$. Here,
the basic idea is that we go to a singular gauge where $\Phi$ is diagonal, and
its eigenvalues $\lambda_{i}$ have branch cuts. The monodromy group corresponds to a
permutation group acting on these five letters. The orbits of the monodromy
group then define the zero modes of the intersecting brane configuration. This
is a convenient characterization, because Yukawa couplings correspond to
combinations of the zero modes which are invariant under $SU(5)_{GUT}$, with
$\lambda_{i}$ assignments which sum to zero. We caution, however,
that the choice of $\Phi$ must be sufficiently generic for this analysis to apply.
The more precise way to proceed is in terms of the T-brane configuration, where
the Yukawa couplings are determined by a residue computation in a gauge in which no
branch cuts are present. We refer to \cite{TBRANES} (see also
\cite{Chiou:2011js, CordovaToAppear}) for further discussion.

A useful first approximation for determining which components of a localized
zero mode can be gauged to zero is as follows. Denote by $V$ the vector space
associated with the $5$ of $SU(5)_{\bot}$ and $\Lambda^{2}V$ the vector space
associated with the $10$ of $SU(5)_{\bot}$. The vector spaces $V$ and
$\Lambda^{2}V$ admit a grading in terms of the generator $T_{3}$, associated
with the diagonal $SU(2)$ defined by the nilpotent $\Phi(0,0)$. We can
therefore write:%
\begin{align}
V  &  =\underset{q}{\oplus}V^{(q)}\\
\Lambda^{2}V  &  =\underset{q}{\oplus}\left(  \Lambda^{2}V\right)  ^{(q)}.
\end{align}
The key point for us is that $\Phi(0,0)$ is a direct sum of nilpotent Jordan
blocks. In other words, acting by $\Phi(0,0)$ on an element of $V$ or
$\Lambda^{2}V$ tends to increase the $T_{3}$ charge. Given our presentation of the
zero modes, we can see that the lowest $T_{3}$ charge vectors
are those which do not have an image under the action of
$\Phi(0,0)$. In other words, we cannot gauge away the lowest $T_{3}$ charge
component of the orbit. Finally, let us note that a full characterization of
localized zero modes requires specifying the position dependence of $\Phi$.

Let us now turn to the couplings between probe D3-branes and T-branes
\cite{TBRANES}. As observed in \cite{FCFT, TBRANES}, most of the details of
$\Phi(Z_{1},Z_{2})$ drop out in the infrared of the probe theory. This is
because in the deformation $\Tr(\Phi\cdot \mathcal{O})$, most of the terms
correspond to irrelevant operator deformations. This means that for the
purposes of specifying the resulting probe theory, a \textquotedblleft
coarse-grained\textquotedblright\ T-brane configuration will suffice
\cite{TBRANES}. The actual localized zero mode content will of course depend
on these higher order terms.

The dominant coupling of the localized zero modes to the probe D3-brane is
determined as follows. First, we go to a gauge where we
eliminate most of the components of the localized $\Psi$-mode. In this gauge,
we can isolate the coupling of the probe to the visible sector mode. In most
cases, there will be a single weight which survives, and as we have already
mentioned, this component of the zero mode will have lowest possible $T_{3}$
charge. The operator~$\mathcal{O}$ which couples to $\Psi$ has conjugate
quantum numbers under $SU(5)_{GUT}\times SU(5)_{\bot}$. In other words, if a
particular component of $R_{\bot}$ for $\Psi$ has $T_{3}$ charge $q$, then the
corresponding $\mathcal{O}$ operator from the D3-brane has $T_{3}$ charge
$-q$. We refer to \cite{TBRANES} for further discussion on this point. Having
presented a general discussion, we now turn to several examples.

\section{Example Scenarios}

In this Appendix we define the coarse-grained T-brane configurations
considered in the main body of the paper, and study the resulting IR fixed
points of the probe theories. Here, our aim is to include only the basic
features of a viable visible sector. To this end, we only require the
existence of matter curves for the $5_{H}$, $\overline{5}_{H}$, $\overline
{5}_{M}$ and $10_{M}$ as well as the interaction terms $5_{H}\times
10_{M}\times10_{M}$ and $\overline{5}_{H}\times\overline{5}_{M}\times10_{M}$.
Additionally, based on flavor considerations in the visible sector, we assume
that all three generations of a particular representation are realized on the
same curve, and that the Yukawas and matter fields all descend from a local
enhancement to $E_{8}$.

The detailed features of the intersection are controlled by a choice of
holomorphic $\Phi$ valued in the $SU(5)_{\bot}$ factor of $SU(5)_{GUT}\times
SU(5)_{\bot}\subset E_{8}$. For the purposes of specifying the infrared
behavior of the probe theory, it is actually enough to expand $\Phi$ to linear
order in the $Z_{i}$, since higher order terms are irrelevant deformations in
the probe theory. For this reason we shall only write out those parts of
$\Phi(Z_{1},Z_{2})$ which survive as relevant and marginal deformations of the
probe theory. This defines a \textquotedblleft
coarse-grained\textquotedblright\ T-brane configuration \cite{TBRANES}. Other than the
requirement that an appropriate Yukawa can be formed, we will not need to
specify the exact profile of matter localization. In what follows, we focus on the probe theory
defined by a single D3-brane. Additional properties of the single D3-brane probe theory, as
well as the case of probes by two D3-branes are summarized in section \ref{sec:opbetafunc}.

\subsection{$\mathbb{Z}_{2}\times\mathbb{Z}_{2}$ Monodromy}

In this section we consider an example of a coarse-grained T-brane
configuration studied in \cite{FCFT,TBRANES} given by:%
\begin{equation}
\Phi=\left[
\begin{array}
[c]{ccccc}%
0 & 1 &  &  & \\
Z_{1} & 0 &  &  & \\
&  & 0 & 1 & \\
&  & Z_{2} & 0 & \\
&  &  &  & 0
\end{array}
\right]  .
\end{equation}
The characteristic polynomial for $\Phi$ has Galois group $\mathbb{Z}%
_{2}\times\mathbb{Z}_{2}$. The generator $T_{3}$ in this case is:%
\begin{equation}
T_{3}=\text{diag}(1/2,-1/2,1/2,-1/2,0).
\end{equation}
Working out the matter curve assignments for the various localized modes, we
consider a coarse-grained T-brane configuration compatible with the $%
\mathbb{Z}
_{2}\times%
\mathbb{Z}
_{2}$ monodromy scenario considered in \cite{EPOINT}. In a singular branched
gauge where $\Phi$ has been diagonalized, the first $%
\mathbb{Z}
_{2}$ factor permutes the eigenvalues $\lambda_{1}$ and $\lambda_{2}$, and the
second permutes the eigenvalues $\lambda_{3}$ and $\lambda_{4}$.
We have that the visible sector modes fill out the following orbits
under the action of the monodromy group:%
\begin{align}
10_{M}  &  :\{\lambda_{1},\lambda_{2}\}\\
5_{H}  &  :\{-\lambda_{1}-\lambda_{2}\}\\
\overline{5}_{H}  &  :\{\lambda_{1}+\lambda_{3},\lambda_{2}+\lambda
_{3},\lambda_{1}+\lambda_{4},\lambda_{2}+\lambda_{4}\}\\
\overline{5}_{M}  &  :\{\lambda_{3}+\lambda_{5},\lambda_{4}+\lambda_{5}\}.
\end{align}
The $T_{3}$ charge assignments for the modes which couple to the probe
D3-brane are:%
\begin{equation}%
\begin{tabular}
[c]{|c|c|c|c|c|}\hline
$%
\mathbb{Z}
_{2}\times%
\mathbb{Z}
_{2}$ & $H_{u}$ & $H_{d}$ & $\overline{5}_{M}$ & $10_{M}$\\\hline
Vector & $e_{1}^{\ast}\wedge e_{2}^{\ast}$ & $e_{2}\wedge e_{4}$ &
$e_{4}\wedge e_{5}$ & $e_{2}$\\\hline
$T_{3}$ charge & $0$ & $-1$ & $-1/2$ & $-1/2$\\\hline
\end{tabular}
\ \ \ \ \ \ .
\end{equation}
The group theory parameters entering $R_{IR}$ are fixed
to be $\mu_{1}=5/2$, $\mu_{2}=5/2$ and $r=2$. In this scenario, all modes
except $H_{u}$ develop a non-zero anomalous dimension. Performing
$a$-maximization, we find $t_{\ast}=0.45$, and the IR scaling dimensions of the
operators are:%
\begin{equation}%
\begin{tabular}
[c]{|c|c|c|c|c|c|c|c|c|c|}\hline
IR dimension & $t_{\ast}$ & $H_{u}$ & $\mathcal{O}_{H_{u}}$ & $H_{d}$ &
$\mathcal{O}_{H_{d}}$ & $\overline{5}_{M}$ & $\mathcal{O}_{\overline{5}_{M}}$
& $10_{M}$ & $\mathcal{O}_{10_{M}}$\\\hline
$%
\mathbb{Z}
_{2}\times%
\mathbb{Z}
_{2}$ & $0.45$ & $1$ & $2.33$ & $1.35$ & $1.65$ & $1.01$ & $1.99$ & $1.01$ &
$1.99$\\\hline
\end{tabular}
\ \ \ \ \ \ \ .
\end{equation}
In other words, the Standard Model chiral matter picks up a small anomalous
dimension, and the Higgs down picks up a slightly larger anomalous dimension.

Finally, let us consider the suppression of the Yukawa couplings for the modes
which have maximal coupling to the probe D3-brane sector:
\begin{equation}
\lambda_{5\times10\times10}  \sim\left(  \frac{M_{\cancel{CFT}}}{M_{GUT}%
}\right)  ^{0.02}\,\,,\,\, \lambda_{\overline{5}\times\overline{5}\times10} \sim\left(
\frac{M_{\cancel{CFT}}}{M_{GUT}}\right)  ^{0.37}.
\end{equation}
For $M_{\cancel{CFT}}/M_{GUT}\sim10^{-3}$, this leads to $\lambda
_{5\times10\times10}\sim0.9$, $\lambda_{\overline{5}\times\overline{5}%
\times10}\sim0.08$.

\subsection{$S_{3}$ Monodromy \label{ExampleS3mono}}

In this section we consider an example based on $S_{3}$ monodromy given by the
coarse-grained T-brane configuration:%
\begin{equation}
\Phi=\left[
\begin{array}
[c]{ccccc}%
0 & 1 & 0 &  & \\
0 & 0 & 1 &  & \\
Z_{1} & Z_{2} & 0 &  & \\
&  &  & 0 & \\
&  &  &  & 0
\end{array}
\right]  .
\end{equation}
The characteristic polynomial for $\Phi$ has Galois group $S_{3}$. Including
higher order terms in $\Phi$, this choice leads to a Dirac neutrino scenario
of the type considered in \cite{BHSV,EPOINT}. In the limit where the Standard
Model fields are non-dynamical, the resulting IR fixed point was studied in
\cite{FCFT}. The $T_{3}$ generator is:%
\begin{equation}
T_{3}=\text{diag}(1,0,-1,0,0).
\end{equation}

Let us now list the various orbits under the action of the monodromy group,
with conventions as in \cite{BHSV}. Since we require the $10_{M}$ to transform
non-trivially, we can go to a singular \textquotedblleft branched
gauge\textquotedblright\ and fix the orbits of the various modes to be:%
\begin{align}
10_{M}  &  :\{\lambda_{1},\lambda_{2},\lambda_{3}\}\\
5_{H}  &  :\{-\lambda_{1}-\lambda_{2},-\lambda_{1}-\lambda_{3},-\lambda
_{2}-\lambda_{3}\}\\
\overline{5}_{M}  &  :\{\lambda_{1}+\lambda_{4},\lambda_{2}+\lambda
_{4},\lambda_{3}+\lambda_{4}\}\\
\overline{5}_{H}  &  :\{\lambda_{1}+\lambda_{5},\lambda_{2}+\lambda
_{5},\lambda_{3}+\lambda_{5}\}
\end{align}
The $T_{3}$ charges and vector assignments for the visible sector modes are:%
\begin{equation}%
\begin{tabular}
[c]{|c|c|c|c|c|}\hline
$S_{3}$ & $H_{u}$ & $H_{d}$ & $\overline{5}_{M}$ & $10_{M}$\\\hline
Vector & $e_{1}^{\ast}\wedge e_{2}^{\ast}$ & $e_{3}\wedge e_{5}$ &
$e_{3}\wedge e_{4}$ & $e_{3}$\\\hline
$T_{3}$ charge & $-1$ & $-1$ & $-1$ & $-1$\\\hline
\end{tabular}
\ \ \ \ \ \ \ .
\end{equation}
Performing $a$-maximization in this case, we note that the group theoretic
parameters entering into $R_{IR}$ are now given as $\mu_{1}=7/2$, $\mu
_{2}=5/2$, $r=4$. The resulting value of $t_{\ast}=0.36$ and the IR scaling
dimensions are:%
\begin{equation}%
\begin{tabular}
[c]{|c|c|c|c|c|c|c|c|c|}\hline
IR dimension & $H_{u}$ & $\mathcal{O}_{H_{u}}$ & $H_{d}$ & $\mathcal{O}%
_{H_{d}}$ & $\overline{5}_{M}$ & $\mathcal{O}_{\overline{5}_{M}}$ & $10_{M}$ &
$\mathcal{O}_{10_{M}}$\\\hline
$S_{3}$ & $1.08$ & $1.92$ & $1.08$ & $1.92$ & $1.08$ & $1.92$ & $1.08$ &
$1.92$\\\hline
\end{tabular}
.
\end{equation}

Let us consider the suppression of the Yukawa couplings associated with the
superpotential couplings~$W_{MSSM}$. For the modes which have maximal coupling
to the probe D3-brane sector, we have:%
\begin{equation}
\lambda_{5\times10\times10}\sim\lambda_{\overline{5}\times\overline{5}%
\times10}\sim\left(  \frac{M_{\cancel{CFT}}}{M_{GUT}}\right)  ^{0.24}%
\end{equation}
For $M_{\cancel{CFT}}/M_{GUT}\sim10^{-3}$, this leads to $\lambda
_{5\times10\times10}\sim\lambda_{\overline{5}\times\overline{5}\times10}%
\sim0.2$.

\subsection{$Dih_{4}$ Monodromy}

Let us now consider an example in which the monodromy group is as large as
possible, while still retaining three distinct $5$ matter curves. We consider
the coarse-grained T-brane configuration:%
\begin{equation}
\Phi=\left[
\begin{array}
[c]{ccccc}%
0 & 1 & 0 & 0 & \\
0 & 0 & 1 & 0 & \\
0 & 0 & 0 & 1 & \\
Z_{1} & 0 & Z_{2} & 0 & \\
&  &  &  & 0
\end{array}
\right]  .
\end{equation}
The characteristic polynomial for $\Phi$ has Galois group $Dih_{4}$. The
generator $T_{3}$ is given by:%
\begin{equation}
T_{3}=\text{diag}(3/2,1/2,-1/2,-3/2,0).
\end{equation}

For concreteness, we focus on the two $Dih_{4}$ monodromy scenarios considered in
\cite{EPOINT}. The analysis of the zero mode content for this scenario
contains a few subtleties, and it is more helpful to
proceed directly to the explicit vector transforming
under $SU(5)_{GUT}$ which cannot be gauged away:
\begin{align}
&
\begin{tabular}
[c]{|c|c|c|c|c|}\hline
$Dih_{4}^{(1)}$ & $H_{u}$ & $H_{d}$ & $\overline{5}_{M}$ & $10_{M}$\\\hline
Vector & $e_{1}^{\ast}\wedge e_{2}^{\ast}$ & $e_{4}\wedge e_{5}$ &
$e_{1}\wedge e_{4},e_{2}\wedge e_{3}$ & $e_{4}$\\\hline
$T_{3}$ charge & $-2$ & $-3/2$ & $0$ & $-3/2$\\\hline
\end{tabular}
\ \ .\\
&
\begin{tabular}
[c]{|c|c|c|c|c|}\hline
$Dih_{4}^{(2)}$ & $H_{u}$ & $H_{d}$ & $\overline{5}_{M}$ & $10_{M}$\\\hline
Vector & $e_{1}^{\ast}\wedge e_{4}^{\ast},e_{2}^{\ast}\wedge
e_{3}^{\ast}$ & $e_{4}\wedge e_{5}$ & $e_{3}\wedge e_{4}$ & $e_{4}$\\\hline
$T_{3}$ charge & $0$ & $-3/2$ & $-2$ & $-3/2$\\\hline
\end{tabular}
\end{align}
The reason for two entries in $H_{u}$ and $\overline{5}_{M}$ is that in a
sufficiently coarse grained T-brane configuration, one or the other of these
modes, but not both, can be gauged away. This subtlety is not particularly
important for us here, because the $T_{3}$ charge assignments for both vectors
are the same.

In this case, the group theoretic parameters entering into $R_{IR}$ are
$\mu_{1}=9/2$, $\mu_{2}=5/2$, $r=10$. Additionally, the Standard Model fields
which couple via a UV relevant operator are those fields with negative $T_{3}$
charge. Performing $a$-maximization, we find the IR scaling dimensions:%
\begin{equation}%
\begin{tabular}
[c]{|c|c|c|c|c|c|c|c|c|c|}\hline
IR dimension & $t_{\ast}$ & $H_{u}$ & $\mathcal{O}_{H_{u}}$ & $H_{d}$ &
$\mathcal{O}_{H_{d}}$ & $\overline{5}_{M}$ & $\mathcal{O}_{\overline{5}_{M}}$
& $10_{M}$ & $\mathcal{O}_{10_{M}}$\\\hline
$Dih_{4}^{(1)}$ & $0.28$ & $1.25$ & $1.75$ & $1.04$ & $1.96$ & $1$ & $2.58$ &
$1.04$ & $1.96$\\\hline
$Dih_{4}^{(2)}$ & $0.27$ & $1$ & $2.59$ & $1.02$ & $1.98$ & $1.22$ & $1.78$ &
$1.02$ & $1.98$\\\hline
\end{tabular}
\end{equation}
Note that in both cases, there is
very little shift to the scaling dimensions of the operators.

We can also see that there is only a mild change in the Yukawa couplings.
Consider the $Dih_{4}^{(2)}$ scenario. For the modes which have maximal
coupling to the probe D3-brane sector, we have:%
\begin{equation}
\lambda_{5\times10\times10}  \sim\left(  \frac{M_{\cancel{CFT}}}{M_{GUT}%
}\right)  ^{0.04}\,\,,\,\, \lambda_{\overline{5}\times\overline{5}\times10} \sim\left(
\frac{M_{\cancel{CFT}}}{M_{GUT}}\right)  ^{0.26}%
\end{equation}
For $M_{\cancel{CFT}}/M_{GUT}\sim10^{-3}$, this leads to $\lambda
_{5\times10\times10}\sim0.8$ and $\lambda_{\overline{5}\times\overline
{5}\times10}\sim0.2$.

\newpage
\bibliographystyle{utphys}
\bibliography{D3susy}

\end{document}